\documentclass[12pt]{article}
\usepackage{amscd,amssymb,amsmath,latexsym,enumerate}
\usepackage[mathscr]{euscript}
\usepackage{epsfig}
\usepackage{fancybox}
\usepackage{verbatim}

\usepackage{color}

\title{Spectral flow  of monopole insertion \\ in topological insulators} 

\author{Alan L. Carey$^1$ and Hermann Schulz-Baldes$^2$
\\
\\
{\small $^1$ Mathematical Sciences Institute, Australian National University, Canberra, Australia}\\
{\small  and the School of Mathematics and Applied Statistics, University of Wollongong, Australia}
\\
{\small $^2$Department Mathematik, Friedrich-Alexander-Universit\"at Erlangen-N\"urnberg, Germany}
}

\date{ }

\newtheorem{theo}{Theorem}

\newtheorem{proposi}{Proposition}

\newtheorem{coro}{Corollary}

\newcommand{\CM}{{\mathbb C}}

\newcommand{\RM}{{\mathbb R}}
\newcommand{\SM}{{\mathbb S}}

\newcommand{\ZM}{{\mathbb Z}}

\newcommand{\Aa}{{\cal A}}

\newcommand{\PP}{{\bf P}}

\newcommand{\EE}{{\bf E}}

\newcommand{\Ff}{{\cal F}}

\newcommand{\Tr}{\mbox{\rm Tr}}

\newcommand{\Mm}{{\cal M}}

\newcommand{\Hh}{{\cal H}}

\newcommand{\one}{{\bf 1}}

\newcommand{\SF}{{\rm Sf}} 
 
\newcommand{\Ch}{{\rm Ch}} 
\newcommand{\Ind}{{\rm Ind}} 
 
\newcommand{\Ran}{{\rm Ran}}

\newcommand{\diag}{{\rm diag}}

\newcommand{\DiracPhase}{G} 
\newcommand{\DiracPhaseEntry}{V} 
\newcommand{\FieldStrength}{F} 
\newcommand{\FieldStrengthForm}{{\cal F}} 
\newcommand{\MonopolTrans}{T} 
\newcommand{\Hardy}{Q}

\begin{document}

\maketitle


\begin{abstract}
Inserting a magnetic flux into a two-dimensional one-particle Hamiltonian leads to a spectral flow through a given gap which is equal to the Chern number of the associated Fermi projection. This paper establishes a generalization to higher even dimension by inserting non-abelian monopoles of the Wu-Yang type. The associated spectral flow is then equal to a higher Chern number. For the study of odd spacial dimensions, a new so-called `chirality flow' is introduced which, for the insertion of a monopole, is then linked to higher winding numbers. This latter fact follows from a new index theorem for the spectral flow between two unitaries which are conjugates of each other by a self-adjoint unitary.
\\
%
\hspace{1.5cm}
\end{abstract}

%


\section{Overview}

The motivation for this study is Laughlin's thought experiment \cite{Lau,AP}. It considers a Landau Hamiltonian describing a two-dimensional electron in a constant magnetic field in which a magnetic flux tube is inserted at some point. This produces supplementary discrete spectrum between Landau levels which flows through a given gap while pushing the flux through. The outcome is that the spectral flow is equal to the Chern number of the Fermi projection below the given gap. While the analysis in \cite{Lau,AP} uses the particular form of the Landau operator, the equality of the spectral flow resulting from a flux insertion and the Chern number is a structural fact which can also be referred to as two-dimensional topological charge pump. In particular, no constant magnetic field and no translation invariance are needed for a non-trivial spectral flow, merely a non-vanishing Chern number. For example, a flux inserted into a disordered, but gapped Haldane model \cite{Hal} leads to a unit spectral flow. These results were established in \cite{DS} for a gapped tight-binding Hamiltonian (based on ideas from \cite{Mac}) and are recalled in Subsection~\ref{sec-LaughReview}, together with some background information on the quantum Hall effect.

\vspace{.15cm}

This paper takes the following perspective on the Laughlin argument: it is a tool to test the topological nature of the ground state of the underlying non-interacting Fermionic system via the insertion of a flux tube. Indeed, a non-vanishing spectral flow during the flux insertion indicates a non-trivial topology of the associated Fermi projection, that is, a non-vanishing Chern number or equivalently a non-vanishing index pairing between the Fermi projection and the two-dimensional Dirac operator. This perspective naturally leads to the question: what replaces the Laughlin argument in spacial dimension different from two? Actually there are similar topological invariants in other dimensions that have played a prominent role in the theory of topological insulators \cite{SRFL,Kit,PS2}. The invariants of interest all stem from the $K$-theory classes of the $d$-dimensional torus. In the complex cases, namely if no symmetries invoking real structures are present (like time-reversal or particle-hole symmetry), all these invariants can be calculated by (non-commutative) differential topological tools as (higher) Chern number and (higher) winding numbers. Amongst all invariants in dimension $d$ there is one called the {\em strong topological invariant}. It is the $K$-theory class of the $d$-torus stemming from the $d$-sphere and is the only invariant whose calculation requires the use of the derivatives in all $d$ spacial dimensions. For even $d$ this is a Chern number, while for odd $d$ and a chiral Hamiltonian it is a winding number. The definition of these objects in a classical differential topological setting is recalled in Subsection~\ref{sec-HalfFlux} below, for the non-commutative version required for the study of disordered systems the reader is referred to \cite{PS2}.  The reason why its precise definition is not relevant is  that the strong invariant is related to an index pairing by an index theorem \cite{Bel,PLB,PS} and only this index pairing will enter in the main results below. As the Chern number in the classical Laughlin argument is the strong invariant, the above question can be reformulated as: what type of flux insertion allows one to calculate the $d$-dimensional strong invariant as a spectral flow?

\vspace{.15cm}

The answer provided below is the following. A non-abelian monopole in the Clifford degrees of freedom of the Dirac operator can be inserted and leads in even spacial dimension to a spectral flow that is equal to the strong invariant given by the Chern number. In odd spacial dimension, the monopole does not lead to a classical spectral flow of the Hamiltonian, but rather a new type of spectral flow that we term a 'chirality flow' which in turn is then given by the strong invariant. In view of these results, the classical Laughlin argument is the special case in dimension $2$. The case of dimension $d=1$ is somewhat special and discussed separately in the introduction to Section~\ref{sec-ChirFlow} which also serves as an elementary illustration of the chirality flow. In higher dimension, the main challenge is the construction of the monopoles and then their insertion in tight-binding lattice translations. The main inspiration for the first point comes from the Wu-Yang construction of a non-abelian monopole in dimension $d=3$ \cite{WY,Shn}. A conceptual approach transposing to arbitrary dimension is developed in  Section~\ref{sec-Monopol}. Moreover, the new monopoles are thoroughly investigated, in particular, their rotational covariance properties and field strength, the associated Yang-Mills field equations as well as their topological charge. For the second point, namely the construction of non-abelian lattice translations, Section~\ref{sec-MonopolTrans} follows the strategy outlined by Arai \cite{Ara} for the construction of magnetic translations. Once these constructions are completed, it is relatively straightforward to insert the monopole in the topological tight-binding Hamiltonian and establish the above claims on the associated spectral and chirality flow (Sections~\ref{sec-SFHam} and \ref{sec-ChirFlow}).

\vspace{.15cm}

Before completing this overview with a somewhat more technical description of the two-dimensional Laughlin argument and the chirality flow, let us stress that this work is not a mere mathematical generalization to higher dimension. Clearly, three-dimensional systems are physically relevant, but also effectively higher dimensional systems can be created in driven systems with one or several driving parameters.  For example, there is a four-dimensional quantum Hall effect \cite{ZH,LSP}. The corresponding four-dimensional strong invariants (so-called second Chern numbers) have been shown to be relevant for magneto-electric effects \cite{LP} as well as for experimental set-ups in certain photonic crystals \cite{ZHG} and in atomic systems \cite{SSP}, and potentially also for non-adiabatic effects \cite{Kol}. 

\subsection{Review of the Laughlin argument for quantum Hall systems}
\label{sec-LaughReview}

Non-interacting quantum Hall systems are two-dimensional Fermionic systems having a topologically quantized Hall conductivity which is equal to the Chern number of the Fermi projection \cite{Bel,ASS,BES,PS2}. The non-trivial topology can result from a constant magnetic field (as in the Landau operator) or a periodic magnetic field (as in the Haldane model \cite{Hal}). Let us describe the relevant mathematical facts for a short-range tight-binding Hamiltonian $h$ on a lattice Hilbert space $\ell^2(\ZM^2)$, or more precisely a covariant family $h=(h_\omega)_{\omega\in\Omega}$ of such operators indexed by a disorder configuration $\omega$ taken from a compact space $\Omega$ on which is given a group action of $\ZM^2$ and an invariant and ergodic probability measure $\PP$ (see the above references for precise discussions on all of these notions which, however, are not essential for the statements in the present paper). If $\mu$ lies in a spectral gap of $h$ and $p=\chi(h\leq \mu)$ is the associated Fermi projection, the associated Chern number is defined as
$$
\Ch(p_\mu)
\;=\;
2\pi\imath\;
\EE_\PP\,
\langle 0|p_\mu[\imath[p_\mu,X_1],\imath[p_\mu,X_2]]|0\rangle
\;,
$$
where $|n\rangle\in\ell^2(\ZM^2)$ is the Dirac notation for a state localized at $n=(n_1,n_2)\in\ZM^2$ and $0\in\ZM^2$ is the origin, and $\EE_\PP$ denotes the average w.r.t. $\PP$ and finally $X_j$ are the two components of the unbounded position operator on $\ell^2(\ZM^2)$ defined by $X_j|n\rangle=n_j|n\rangle$ for $j=1,2$. The Kubo formula shows that the zero-temperature Hall conductance is equal to $\Ch(p)$ \cite{Bel,BES}. An index theorem furthermore shows that
\begin{equation}
\label{eq-2dInd}
\Ch(p_\mu)
\;=\;
\Ind(p_\mu Vp_\mu)
\;,
\qquad
V\;=\;
\frac{X_1+\imath X_2}{|X_1+\imath X_2|}
\;,
\end{equation}
where $p_\mu Vp_\mu$ is a Fredholm operator on $\Ran(p_\mu)$ and the index on the r.h.s. is known to be almost surely constant. For reasons explained further below, $V$ is also called the Dirac phase. While the Chern number is only defined for covariant Hamiltonians, the index $\Ind(p_\mu Vp_\mu)$ in \eqref{eq-2dInd} is defined for any local Hamiltonian with a gap at $\mu$. Locality means by definition that $\langle n|h|m\rangle$ decays sufficiently fast in $|n-m|$. The Laughlin argument described next only uses this index and is hence a purely spectral-theoretic statement about one fixed Hamiltonian. Adapting Laughlin's idea one adds a magnetic flux $\alpha\in\RM$ to just one specified cell of the lattice $\ZM^2$. This can be done by a rotationally symmetric gauge potential (for details see \cite{DS} or Section~\ref{sec-MonopolTrans} below). This results in a one parameter family $\alpha\in\RM\mapsto h_\alpha$ of bounded local Hamiltonians on $\ell^2(\ZM^2)$ with $h_0=h$. The main results of \cite{DS} are then:
\begin{itemize}
\item[{\rm (i)}] $h_\alpha-h_0$ is a compact operator, so that $h_\alpha$ and $h_0$ have the same essential spectrum.

\item[{\rm (ii)}] $h_{\alpha+1}=\DiracPhaseEntry ^*h_\alpha \DiracPhaseEntry $ where $\DiracPhaseEntry$ is as in \eqref{eq-2dInd}.

\item[{\rm (iii)}]  The spectral flow $\SF(\alpha\in[0,1]\mapsto h_\alpha-\mu)$ is equal to $\Ind(p_\mu Vp_\mu)$.

\end{itemize}

Items (i) and (ii) are linked to the rotationally symmetric gauge. For other choices of the gauge, one may not have compactness of $h_\alpha-h_0$ even though $h_\alpha$ and $h_0$ still have the same spectrum and the spectral flow is the same. The particular relation $h_{\alpha+1}=\DiracPhaseEntry^*h_\alpha \DiracPhaseEntry $ in (ii) is not essential, crucial is merely the norm continuity of $\alpha\to h_\alpha$ and that the initial point $h=h_0$ and final point $h_1$ are unitarily equivalent. Item (iii) is the main result of \cite{DS} and the proof is essentially an application of  Phillips' results connecting spectral flow to an index (herein Theorem~\ref{theo-SFInd1} recalled for the readers' convenience in Appendix~\ref{sec-SF}). Let us stress that $h_\alpha$ is {\it not} equal to $(\DiracPhaseEntry^\alpha)^* h_0 \DiracPhaseEntry^\alpha$ where $\DiracPhaseEntry^\alpha$ is the $\alpha$-th root. Such a unitary equivalence would imply that there is no spectral flow. 

\subsection{Generalization to higher even dimensions}

In higher even dimension $d$ the strong invariant is given by the higher Chern number $\Ch_d(p_\mu)\in\ZM$ where $p_\mu=\chi(h\leq \mu)$ is still the Fermi projection of a local possibly matrix-valued Hamiltonian $h$ on $\ell^2(\ZM^d,\CM^n)$ below a gap $\mu$. The definition of the higher Chern number given in \cite{PLB,PS2} is irrelevant for the present purposes as $\Ch_d(p_\mu)$ is again linked to the index of a Fredholm operator by an index theorem completely analogously to \eqref{eq-2dInd}, provided the unitary operator $V$ is chosen as follows. The $d$-dimensional (dual) Dirac operator is given by
\begin{equation}
\label{eq-Ddef}
D\;=\;\sum_{j=1}^d\gamma_j X_j
\;,
\end{equation}
where $X_j$ are the $d$ components of the position operator and $\gamma_1,\ldots,\gamma_d$ is an irreducible representation of the Clifford algebra acting on an auxiliary finite dimensional representation space $\CM^{2N}$ (see \cite{LM,PS2} or Section~\ref{sec-MonopolePot} for details on this representation). Note that upon discrete Fourier transform, $D$ becomes the standard first order Dirac operator on the $d$-dimensional torus. Associated to the Dirac operator is also its selfadjoint Dirac phase $G$ by
\begin{equation}
\label{eq-DGdef}
\DiracPhase \;=\;D|D|^{-1}
\;,
\end{equation}
In even dimension $d$, the Dirac operator has a chiral symmetry $\Gamma D\Gamma=-D$ for a suitable selfadjoint unitary matrix $\Gamma$ acting on the representation space. In the spectral representation of $\Gamma$ one has  $\Gamma=\diag(\one_N,-\one_N)$ and therefore there is a unitary $V$ such that the Dirac phase is of the form
\begin{equation}
\label{eq-Vdef}
G
\;=\;
\begin{pmatrix}
0 & V^* \\ V & 0
\end{pmatrix}
\;.
\end{equation}
Note that for $d=2$, the matrices $\gamma_1,\gamma_2,\Gamma$ are given by the standard Pauli matrices and therefore $V$ is as in \eqref{eq-2dInd}. Now with $V$ from \eqref{eq-Vdef} and $P_\mu=p_\mu\otimes \one$,  an index theorem similar to that in \eqref{eq-2dInd} still connects $\Ch_d(p_\mu)$ to $\Ind(P_\mu VP_\mu)$ \cite{PLB,PS2}. Again the index  $\Ind(P_\mu VP_\mu)$ makes sense for any local Hamiltonian $h$, and this is the only data needed for the higher dimensional argument.

\vspace{.15cm}

The first major result of this paper (Theorem~\ref{theo-SpecFlow} in Section~\ref{sec-SFHam}) states that items (i), a modified version of (ii) and (iii) in Subsection~\ref{sec-LaughReview} also hold for matrix-valued Hamiltonians in higher even dimensions, provided that  the path $\alpha\in\RM\mapsto h_\alpha$ is obtained by inserting the higher dimensional analogue of a non-abelian Wu-Yang monopole \cite{WY,Shn}.  The construction of these monopoles and consequently their insertion in a lattice Hamiltonian take up a large part of the paper, notably Sections~\ref{sec-Monopol} and \ref{sec-MonopolTrans}. Let us stress a major difference between the two and higher dimensional case: the monopoles in dimension $d>2$ satisfy the Yang-Mills equation only for $\alpha\in\{0,\frac{1}{2},1\}$, see Proposition~\ref{prop-YM}.

\subsection{Chirality flow for systems in odd dimension}
\label{sec-ChiralFlow}

The second major result of the paper (Theorem~\ref{theo-ChirFlow} in Section~\ref{sec-ChirFlow}) shows what replaces the Laughlin argument for chiral local Hamiltonians $h$ in odd space dimension $d$. If $h$ acts on $\ell^2(\ZM^d,\CM^{2n'})$, that is $n=2n'$ is even, the chiral symmetry operator is  $J=\mbox{\rm diag}(\one_{n'},-\one_{n'})$ and the the chiral symmetry of $h$ reads
$
J\,h\,J
\;=\;-h
\;.
$
The Fermi level in such systems is $\mu=0$ so that $h$ is invertible and the so-called flat band Hamiltonian is of the form
\begin{equation}
\label{eq-FermiUnitary}
h |h|^{-1}
\;=\;
\begin{pmatrix}
0 & u^* \\ u & 0
\end{pmatrix}
\;,
\end{equation}
for a unitary operator $u$ on $\ell^2(\ZM^d,\CM^{n'})$. This unitary is called the Fermi unitary \cite{PS2} because it uniquely determines the Fermi projection $p_\mu=\frac{1}{2}(\one-h |h|^{-1})$ of the chiral Hamiltonian. The Fermi unitary has a $d$-dimensional winding number $\Ch_d(u)\in\ZM$ as a strong invariant which is also called an odd Chern number \cite{PS}. Again the reader is referred to \cite{PS,PS2} for the definition because all that is relevant in the following is the link to the index of a Fredholm operator (Corollary 6.3.2 in \cite{PS2}):
$$
\Ch_d(u)
\;=\;-\,
\Ind(\Hardy U\Hardy)
\;,
$$
where $\Hardy=\chi(D>0)=\chi(G>0)=\frac{1}{2}(\DiracPhase +\one)$ and $U=u\otimes\one_{2N}$. While $\Ch_d(u)$ is only defined for a covariant Hamiltonian, the index $\Ind(\Hardy\,U\,\Hardy)$ makes sense for every single chiral local Hamiltonian. The new contribution of the present paper is to calculate this index and thus the strong invariant as a suitably defined spectral flow. For this purpose, one inserts again a non-abelian monopole which in dimension $d=3$ is precisely the Wu-Yang monopole. This provides a path $\alpha\in[0,1]\mapsto H_\alpha$ of chiral Hamiltonians on $\ell^2(\ZM^d,\CM^{2n'})\otimes\CM^{2N}$ with $H_0=h\otimes\one_{2N}$. Generically, this path is invertible so that via \eqref{eq-FermiUnitary}  there are associated Fermi unitaries $\alpha\in[0,1]\mapsto U_\alpha$ with $U_0=U$. Typically, the spectrum of these unitaries fills the whole unit circle. The crucial facts, corresponding to those in the even dimensional case, are:
\begin{itemize}
\item[{\rm (i)$^\prime$}] $U_\alpha-U_0$ is compact.
\item[{\rm (ii)$^\prime$}] $U_1=\DiracPhase U_0\DiracPhase $ where $\DiracPhase =D|D|^{-1}$.
\item[{\rm (iii)$^\prime$}]  $\SF(\alpha\in[0,1]\mapsto \DiracPhase U_\alpha U^*_0)=\Ind(\Hardy U\Hardy)$.
\end{itemize}
\noindent Items (i)$^\prime$ and (ii)$^\prime$ follow again from the construction of the monopole. For $d\geq 3$, there is also a relation $U_\alpha=\DiracPhase U_{1-\alpha}\DiracPhase $ which corresponds to the relation in (ii), but this is of no importance for the definition of the spectral flow and the claim in (iii)$^\prime$. Indeed, the path $\alpha\in[0,1]\mapsto \DiracPhase U_\alpha U^*_0$ of unitaries connects two selfadjoint unitaries $\DiracPhase $ and $U_0\DiracPhase U_0^*$ with spectrum $\{-1,1\}$ and, as also  $\DiracPhase U_\alpha U^*_0 -\DiracPhase $ is compact, the above spectral flow counts the eigenvalues moving between them. Theorem~\ref{theo-SFInd2} in the Appendix allows to show that this spectral flow is equal to the index in (iii)$^\prime$. A precise statement of (iii)$^\prime$ is given in Theorem~\ref{theo-ChirFlow} in Section~\ref{sec-ChirFlow}. It is shows that the index in (iii)' is equal to the spectral flow from $J\DiracPhase $ to $-HJ\DiracPhase H^{-1}$ which justifies the terminology chiral flow. Let us also advertise that the introduction to Section~\ref{sec-ChirFlow} describes a one-dimensional version of this (insertion of a flux in the Su-Schrieffer-Heeger model).

\subsection{Organization of the paper and omissions}

As already stressed above, a large part of the paper is devoted to the construction and analysis of non-abelian monopoles (Section~\ref{sec-Monopol}) and the associated non-abelian monopole translations on the lattice (Section~\ref{sec-MonopolTrans}). Based on this, the higher dimensional Laughlin argument is then proved in Section~\ref{sec-SFHam} for even $d$ and in Section~\ref{sec-ChirFlow} for odd $d$. The latter section uses the notion of chirality flow which is introduced and studied in the Appendix.  Sections~\ref{sec-Monopol} and \ref{sec-MonopolTrans} as well as the Appendix  can be read independently of the remainder of the paper.

\vspace{.15cm}

Let us add a short comment on what is omitted in this paper, but will be dealt with elsewhere. As shown in \cite{DS} for the two-dimensional case and in \cite{CPS} for a particular one-dimensional model, one can implement real symmetries to the Hamiltonian $h_0$ or $H_0$ and then analyse the resulting symmetry properties of the spectral flow when a flux is inserted. These real symmetries are typically given by a time-reversal or a particle-hole symmetry. Such a symmetry analysis is also possible for higher dimensional models. On the index side of the equalities in (iii) and (iii)$^\prime$, this was already carried out in \cite{GS}. When the monopole is inserted, the equation $H_\alpha=\DiracPhase  H_{1-\alpha} \DiracPhase $ will then be relevant. In some situations, one can then prove the existence of bound states for half-flux $\alpha=\frac{1}{2}$, see the example in \cite{DS}.

\section{Non-abelian monopoles}
\label{sec-Monopol}

This section presents a conceptional approach to the construction of static monopoles in classical non-abelian field theory over $\RM^d$ with structure group SU$(N)$. Characteristic properties of such monopoles are a singularity at the origin of $\RM^d$ and a decaying field strength which has a rotational covariance property. For a particular value of the parameters  (notably for half-flux), the monopole satisfies the Yang-Mills field equation and has a topological charge equal to $1$. All these algebraic facts are proved in subsections below. In dimension $d=3$ and for half-flux, these monopoles reduce precisely to the Wu-Yang monopoles \cite{WY}, see Subsection~\ref{sec-Monopole234D}. Historically, Wu and Yang exhibited explicit solutions of the SU$(2)$ Yang Mills equations and their motivation came from the theory of isospin. More recently the Wu-Yang monopole has resurfaced in connection with the fractional Hall effect and a discussion with references is found in \cite{F}. Further information on the Wu-Yang monopole can be found in the monograph \cite{Shn}. Let us stress though that the constructions below work in arbitrary dimension. Interestingly,  in even dimension the monopole acquires a supplementary chiral symmetry. This section is written such that it can be read independently of the rest of the paper and may be of interest in the context of classical non-abelian field theory. The covariant derivatives are analyzed as purely algebraic objects here and further functional analytic issues are then dealt with in Section~\ref{sec-MonopolTrans}. Throughout this section, we suppose that $d\geq 2$ as the following formulas are not interesting in the case $d=1$ for which a separate treatment is given later.
\vspace{.15cm}

The following has some similarities to the construction of instantons in even space dimension as presented by Fujii \cite{Fuj}. This will briefly be discussed at the end of Subsection~\ref{sec-MonopolePot}.

\subsection{General set-up for classical non-abelian gauge theory}
\label{sec-SetupGauge}

To fix our notations, let us describe the framework and collect a few basic general facts on classical non-abelian gauge theory on $\RM^d$ or  a subset of $\RM^d$. We restrict to a SU$(N)$-gauge theory with no external sources (not coupled to any other field so that one speaks of a pure gauge theory). The basic object is a non-abelian gauge potential $A=(A_1,\ldots,A_d)$ which is a collection of functions $x\in\RM^d\mapsto A_k(x)$ in the Lie algebra $\mbox{su}(N)$, namely 
\begin{equation}
\label{eq-ASkew}
A_k(x)^*\;=\;-\,A_k(x)
\;,
\qquad
\Tr(A_k(x))\;=\;0
\;,
\qquad
k=1,\ldots,d
\;.
\end{equation}
Below a particular gauge potential describing a monopole will be introduced. It will not be defined on all $\RM^d$, but only for $x\not=0$.  Given $A$, one next constructs covariant derivatives
$$
\nabla_k
\;=\;
\partial_k\,+\,A_k(x)
\;,
\qquad
k=1,\ldots,d\;.
$$
The non-abelian field strength tensor associated to $A$ is defined as the collection of operators 
$$
\FieldStrength_{k,l}(x)
\;=\;
[\nabla_k,\nabla_l]
\;,
\qquad
k,l=1,\ldots,d
\;.
$$
Actually, these operators are multiplication operators and can also viewed as matrix-valued functions on $\RM^d$. They are given by the standard expression
\begin{equation}
\label{eq-FieldStrengthDef}
\FieldStrength_{k,l}(x)
\;=\;
\partial_k\,A_l(x)\;-\;\partial_l\,A_k(x)
\;+\;
[A_k(x),A_l(x)]
\;.
\end{equation}
Also in the present context the field strength is skewadjoint and antisymmetric in the tensor indices:
$$
\FieldStrength_{k,l}(x)^*
\;=\;
-\,
\FieldStrength_{k,l}(x)
\;,
\qquad
\FieldStrength_{k,l}(x)
\;=\;-\,\FieldStrength_{l,k}(x)
\;.
$$
In particular, the field strength takes values in the Lie algebra su$(N)$. The entries $B_m=\frac{1}{2}\epsilon_{m,k,l}\FieldStrength_{k,l}(x)$ of the field strength are called the non-abelian magnetic fields (here $\epsilon$ is the completely antisymmetric tensor and the Einstein sum convention is used). The field strength is defined as a commutator of covariant derivatives. As these covariant derivatives satisfy a Jacobi identity, this leads to an equation for the field strength which called the second Bianchi identity:
$$
[\nabla_n,\FieldStrength_{k,l}]\,+\,
[\nabla_k,\FieldStrength_{l,n}]\,+\,
[\nabla_l,\FieldStrength_{n,k}]
\;=\;
0
\;.
$$
Another set of equations that, however, a given field strength may or may not satisfy, are the Yang-Mills field equations (in absence of exterior currents). They read $\sum_{k=1}^d[\nabla_k,F_{k,l}]=0$ for $l=1,\ldots,d$, or equivalently
\begin{equation}
\label{eq-YM}
\sum_{k=1}^d
\big(\partial_k F_{k,l}
\;+\;
[A_k,F_{k,l}]\big)
\;=\;0
\;,
\qquad
l=1,\ldots,d
\;.
\end{equation}

\vspace{.15cm}

It will also be useful to slightly generalize the notations. For $v=(v_1,\ldots,v_d)\in\RM^d$ and another matrix-valued function $x\in\RM^d\mapsto A_v(x)$, directional derivatives are introduced by
$$
A_v(x)\;=\;\sum_{k=1}^d v_k A_k(x)
\;=\;
\langle v|A(x)\rangle
\;,
\qquad
\partial_v
\;=\;
\langle v|\partial\rangle
\;,
$$
where here $\langle\,.\,|\,.\,\rangle$ denotes the euclidean scalar product in $\RM^d$, $A(x)=(A_1(x),\ldots,A_d(x))$ and $\partial=(\partial_1,\ldots,\partial_d)$. If $\nabla=(\nabla_1,\ldots,\nabla_d)$ and $\nabla_v=\langle v|\nabla\rangle$, one then also has
$
\nabla_v
\;=\;
\partial_v\,+\,A_v
\;.
$
Furthermore, setting $\FieldStrength_{v,w}(x)=[\nabla_v,\nabla_w]$ for $v,w\in\RM^{d}$, one has
$$
\FieldStrength_{v,w}
\;=\;
\partial_v A_w\,-\,\partial_v A_w\,+\,[A_v,A_w]
\;=\;
-\,\FieldStrength_{w,v}
\;.
$$
Finally let us associate differential forms to the gauge potential and field strength.
The connection $1$-form is a matrix-valued differential form $\Aa$ on $\RM^d$ and the associated field strength is a matrix-valued curvature $2$-form $\Ff$ on $\RM^d$. They are given by
\begin{equation}
\label{eq-ConnForm}
\Aa
\;=\; 
\sum_{k=1}^d\,A_k\,{\rm d}x_k
\;,
\qquad
\FieldStrengthForm
\;=\;
\frac{1}{2}\,
\sum_{k,l=1}^d
\,
\FieldStrength_{k,l}(x)
\;{\rm d}x_k\wedge {\rm d}x_l
\;,
\end{equation}
where the roman ${\rm d}$ denotes exterior differentiation. The Cartan structure equation linking these forms $\FieldStrengthForm=({\rm d}\,\Aa)\;+\;\Aa\wedge\Aa$, and also the Bianchi identity and Yang-Mills equation can be written out using these forms.

\subsection{Construction of the monopole potential}
\label{sec-MonopolePot}

Let $\gamma_1,\ldots,\gamma_d$ be a faithful selfadjoint representation of the $d$ generators of the complex $d$-dimensional Clifford algebra $\CM_d$ on the spinor space $\CM^{2N}$, namely $\gamma_i\gamma_j=-\gamma_j\gamma_i$ for $i\not=j$ and $\gamma_j^*=\gamma_j$ and $(\gamma_j)^2=\one_{2N}$. This representation  can be chosen irreducible if $2N=2^{\frac{d-1}{2}}$ for odd $d$ and $2N=2^{\frac{d}{2}}$ for even $d$. Irreducibility is important for some, but not all results below. If $d$ is even, then there exists a grading operator $\Gamma=(-\imath)^{\frac{d}{2}}\gamma_1\cdots\gamma_d$ on $\CM^{2N}$ with $\Gamma^2=\one$ and $\Gamma^*=\Gamma$ such that $\Gamma\gamma_j=-\gamma_j\Gamma$. For $d$ odd, the representation is chosen such that $\gamma_1\cdots\gamma_d=\imath^{\frac{d-1}{2}}\,\one$. Now introduce the (dual) Dirac operator $D$ defined by \eqref{eq-Ddef} as an unbounded, selfadjoint multiplication operator on $L^2(\RM^d,\CM^{2N})$ with unitary, selfadjoint phase $\DiracPhase =D|D|^{-1}$, see \eqref{eq-DGdef}. As $D$ vanishes at the origin, one has to take care with the definition of $\DiracPhase$. One first defines $\DiracPhase$ on $C^\infty(\RM^d\setminus\{0\},\CM^{2N})$, and then readily checks that the range of $\DiracPhase\pm\imath\,\one$ is dense. Hence $\DiracPhase$ is essentially selfadjoint and similarly one can check that  $\DiracPhase^2=\one$. Upon Fourier transform, $D$ becomes a first-order differential Dirac operator on $L^2(\RM^d,\CM^{2N})$. Later on, we will also consider $D$ as an operator on $\ell^2(\ZM^d,\CM^{2N})$ and then the Fourier transform is the Dirac operator on the $d$-torus which has compact resolvent. Let us further note that $\DiracPhase(x) $ is only defined away from the origin and thus one has to choose $\DiracPhase(0)$ on the lattice Hilbert space $\ell^2(\ZM^d,\CM^{2N})$, see Proposition~\ref{prop-SalphaDef} below. If $d$ is even, 
\begin{equation}
\label{eq-DiracChir}
\Gamma D\Gamma\;=\;-D
\;.
\end{equation}
In the spectral representation of $\Gamma$, this leads to, still only for $d$ even,
\begin{equation}
\label{eq-GammaSpec}
\Gamma
\;=\;
\begin{pmatrix}
\one & 0 \\ 0 & -\one
\end{pmatrix}
\;,
\qquad
\DiracPhase 
\;=\;
\begin{pmatrix}
0 & \DiracPhaseEntry^* \\ \DiracPhaseEntry  & 0
\end{pmatrix}
\;,
\end{equation}
for a unitary $\DiracPhaseEntry $ which can also be expressed in terms of a lower-dimensional representation of $\CM_{d-1}$.  Now the monopole potential with flux $\alpha\in\RM$ is introduced by
\begin{equation}
\label{eq-MonopoleDef}
A^\alpha_k(x)
\;=\;
\alpha\,\DiracPhase ^*\,(\partial_k\,\DiracPhase )
\;,
\qquad
k=1,\ldots,d\;,
\end{equation}
where $\partial_1,\ldots,\partial_d$ are the partial derivatives (diagonally extended, namely $\partial_j$ is identified with $\partial_j\otimes\one_{2N}$). Here the argument $x$ in $A^\alpha_k(x)$ indicates that we consider $A^\alpha_k(x)$ as a matrix-valued function on $\RM^d\setminus\{0\}$, but let us stress that it is also viewed as a multiplication operator on $L^2(\RM^d,\CM^{2N})$ and as such part of the covariant derivatives. Indeed, $\DiracPhase ^*\,(\partial_k\,\DiracPhase )$ can also be written as $\DiracPhase [\partial_k\,\DiracPhase ]$ and hence the definition \eqref{eq-MonopoleDef} can also be written in terms of the associated $1$-form $\Aa^\alpha$ as
$$ 
\Aa^\alpha\;=\;\alpha\,\DiracPhase ^*\,({\rm d}\,\DiracPhase )
\;.
$$
%
Let us also stress that $A^\alpha_k(x)$ is a pure gauge only for $\alpha=0$ and $\alpha=1$ (recall that a vector potential is called a pure gauge if it is a logarithmic derivative of a gauge transformation, that is, a unitary multiplication operator). As in Section~\ref{sec-SetupGauge}, we also use the notation $A^\alpha_v(x)=\langle v|A^\alpha(x)\rangle$. By construction, the monopole potential is skew-adjoint and traceless, that is satisfies \eqref{eq-ASkew} and thus lies in the Lie algebra su$(N)$ of SU$(N)$. Note that for even $d$, one has 
$
\Gamma\, A^\alpha_k\, \Gamma
\;=\;
A_k^\alpha
\;,
$ 
so that $A_k$ is diagonal in the spectral representation \eqref{eq-GammaSpec} of $\Gamma$.  

\vspace{.15cm}

Next let us derive alternative expressions for the monopole potential. This will use the radial operator
$$
R
\;=\;
(X_1^2+\ldots+X_d^2)^{\frac{1}{2}}\;=\;|D|
\;,
$$ 
which is also a diagonal selfadjoint operator on $L^2(\RM^d,\CM^{2N})$.  As $\{D,\gamma_k\}=D\gamma_k+\gamma_kD=2X_k$, one has
$$
\DiracPhase [\partial_k,\DiracPhase ]
\;=\; \DiracPhase \left(\frac{\gamma_k}{R}-X_k\frac{D}{R^3}\right)
\;=\; \left(\frac{D\gamma_k}{R^2}-X_k\frac{D^2}{R^4}\right)
\;=\; \frac{D\gamma_k-X_k}{R^2}
\;=\; \frac{1}{2}\,\frac{[D,\gamma_k]}{R^2}
\;,
$$
which leads to
\begin{equation}
\label{eq-AFormula}
A^\alpha_k(x)
\;=\;
\frac{\alpha}{2}\,\frac{[D,\gamma_k]}{R^2}
\;=\;
\alpha\,\frac{D\gamma_k-x_k}{R^2}
\;=\;
\alpha\,\sum_{j\not=k}\gamma_j\gamma_k\;\frac{x_j}{R^2}
\;.
\end{equation}
Furthermore, the monopole potential is divergence free away from the origin $x=0$:
$$
\sum_{k=1}^d
\partial_k\,A^\alpha_k(x)
\;=\;
\frac{\alpha}{2}\,
\sum_{k=1}^d
\left(
\frac{[\gamma_k,\gamma_k]}{R^2}
\;-\;
2\,\frac{[D,\gamma_k]}{R^3}\,(2\,x_k)
\right)
\;=\;
0
\;.
$$
The monopole potential  is not rotationally invariant, but has a covariance property w.r.t. rotations and the spin representation, see \eqref{eq-SpinCov} below.

\vspace{.15cm}

As in Section~\ref{sec-SetupGauge}, we similarly introduce the notation $\gamma=(\gamma_1,\ldots,\gamma_d)$ and $\gamma_v=\langle v|\gamma\rangle$. Then $\gamma_v$ is a symmetry if $\|v\|=1$:
$$
\gamma_v^*\;=\;\gamma_v\;,
\qquad
\gamma_v^2\;=\;\one
\;.
$$
The group $\mbox{\rm Pin}(d)\subset\CM_d$ is multiplicatively generated by the $\gamma_v$ with $\|v\|=1$ \cite{LM}. Furthermore, the even elements in $\mbox{\rm Pin}(d)$, namely those formed of an even number of $\gamma$'s, form the subgroup $\mbox{\rm Spin}(d)$ \cite{LM}.   Moreover, this notation allows to write 
\begin{equation}
\label{eq-AFormula2}
D=\langle X|\gamma\rangle\;=\;\gamma_X
\;,
\qquad
A^\alpha_v(x)
\;=\;
\frac{\alpha}{2}\;\frac{[\langle x|\gamma\rangle,\langle v|\gamma\rangle]}{R^2}
\;=\;
\frac{\alpha}{2}\;\frac{[\gamma_x,\gamma_v]}{R^2}
\;.
\end{equation}
A crucial property of the monopole potential is its radial scaling property:
\begin{equation}
\label{eq-AFormula3}
A^\alpha_v(rx)
\;=\;
\frac{1}{r}\,A^\alpha_v(x)
\;,
\qquad
r>0
\;.
\end{equation}
In particular, it is completely determined by its values on the unit sphere $\SM^{d-1}\subset\RM^d$. Furthermore, $A_v(x)$ decays (in matrix operator norm) at infinity as 
\begin{equation}
\label{eq-ABound}
\|A^\alpha_v(x)\|\;\leq\;\frac{|\alpha|}{R}
\;.
\end{equation}

Let us conclude this section with a brief comment on the construction of instantons. These are everywhere defined (and smooth) gauge fields which decay and are of pure gauge type at infinity. Instantons also have a topological charge similar to the half-flux monopoles \eqref{eq-MonopoleDef}. However, the latter have a singularity and are not of pure gauge at infinity. In the present context, the construction of the gauge potential $A^{\mbox{\rm\tiny inst}}$ for instantons as given in \cite{Fuj} can be restated as
$$
A^{\mbox{\rm\tiny inst}}_k(x)\;=\;\frac{x^2}{x^2+1}\;G(x)^*\partial_k G(x)
\;,
\qquad
k=1,\ldots,d
\;.
$$
For $d=4$, this is the BPST instanton and in higher dimension it is connected to \cite{Fuj}. The topological charge (also called Pontryagin number or instanton number) can be calculated by similar techniques as in the subsections below.

\subsection{Covariant derivatives of a monopole}
\label{sec-CovDeriv}

The covariant derivative in the field of a monopole with flux $\alpha\in\RM$ is defined by
\begin{equation}
\label{eq-CovDer}
\nabla_k^\alpha
\;=\;
\partial_k\,+\,A^\alpha_k(x)
\;,
\qquad
k=1,\ldots,d\;.
\end{equation}
As in Section~\ref{sec-SetupGauge}, we then also use the notations $\nabla^\alpha=(\nabla^\alpha_1,\ldots,\nabla^\alpha_d)$ and $\nabla^\alpha_v=\langle v|\nabla^\alpha\rangle$ for  $v\in\RM^d$. As the $A^\alpha_k(x)$ are skew-adjoint, the operators $\imath\nabla_v^\alpha$ are formally selfadjoint operators and it will be shown in Section~\ref{sec-Selfadjointness} that they actually define selfadjoint operators on $L^2(\RM^d,\CM^{2N})$.  An important algebraic relation is
\begin{equation}
\label{eq-NablaConj}
\DiracPhase \,\nabla_v^\alpha\,\DiracPhase 
\;=\;
\partial_v
\;+\;
\DiracPhase [\partial_v,\DiracPhase ]
\;+\;
\DiracPhase  A^\alpha_v(x)\DiracPhase 
\;=\;\nabla_v^{1-\alpha}
\;.
\end{equation}
This follows from \eqref{eq-AFormula} combined with
$$
\DiracPhase \,\frac{1}{2}\,\frac{[D,\gamma_k]}{R^2}\, \DiracPhase 
\;=\;
\frac{D}{R}\,\frac{1}{2}\,\frac{[D,\gamma_k]}{R^2}\, \frac{D}{R}
\;=\;
\frac{1}{2R^4}
\;D(D\gamma_k-\gamma_kD) D
\;=\; 
-\,\frac{1}{2}\,\frac{[D,\gamma_k]}{R^2}
\;.
$$
Let us also note that for even $d$, one has $[\nabla_k^\alpha,\Gamma]=0$ so that $\nabla_k^\alpha$ is diagonal in the spectral representation of $\Gamma$.

\subsection{Field strength and Yang-Mills equations of a monopole}
\label{sec-FieldStrength}

Associated to $A^\alpha$ one now has via \eqref{eq-FieldStrengthDef} the non-abelian field strength tensor $\FieldStrength^\alpha_{k,l}$ of the $\alpha$-monopole. With some care, one can calculate the field strength explicitly from \eqref{eq-AFormula}, namely from
$$
\partial_k\,A^\alpha_l(x)
\;=\;
\frac{\alpha}{R^4}\,
\Big(
R^2\,\gamma_k\gamma_l\;-\;\sum_{j\not=l}2\,x_jx_k\,\gamma_j\gamma_l
\Big)
$$
and
$$
[A^\alpha_k(x),A^\alpha_l(x)]
\;=\;
\frac{2\alpha^2}{R^4}\,
\Big(-\,
R^2\,\gamma_k\gamma_l\;-\;\sum_{j\not=k}x_jx_k\,\gamma_j\gamma_l
\;+\;\sum_{j\not=l}x_jx_l\,\gamma_j\gamma_k
\Big)
\;,
$$
one deduces
\begin{align}
\FieldStrength^\alpha_{k,l}(x)
& 
\;=\;
\frac{2(\alpha^2-\alpha)}{R^4}\,
\sum_{j=1,\ldots,d}\Big(x_j^2\,
\gamma_l\gamma_k
\;-\;
x_jx_l\, \gamma_j\gamma_k
\;+\;
x_jx_k\, \gamma_j\gamma_l
\Big)
\nonumber
\\
&
\;=\;
\frac{2(\alpha^2-\alpha)}{R^4}\,
\big(R^2\gamma_l\gamma_k
\;+\;
D(x_k\gamma_l-x_l\gamma_k)\big)
\;.
\label{eq-MonopoleField}
\end{align}
Therefore for $\alpha=0$ and $\alpha=1$ the field strength vanishes, in accordance with the fact noted above that $A^0$ and $A^1$ are pure gauge fields. Furthermore, in the same way as in \eqref{eq-AFormula3}, one has
\begin{equation}
\label{eq-FieldStrengthScale}
\FieldStrength^\alpha_{k,l}(rx)
\;=\;
\frac{1}{r^2}\,
\FieldStrength^\alpha_{k,l}(x)
\;,
\qquad
r>0
\;,
\end{equation}
and thus
$$
\|\FieldStrength^\alpha_{k,l}(x)\|\;\leq\;\frac{6\,|\alpha^2-\alpha|}{R^2}
\;.
$$
For even dimension $d$, 
$
\Gamma\,\FieldStrength^\alpha_{k,l}(x)\,\Gamma
\;=\;
\FieldStrength^\alpha_{k,l}(x)
\;.
$
Both sums in the Yang-Mills field equations \eqref{eq-YM} can be calculated explicitly for the monopole gauge potential $A^\alpha$:

\begin{proposi}
\label{prop-YM}
One has
$$
\sum_{k=1}^d
\partial_k F_{k,l}^\alpha
\;=\;
\frac{2(\alpha^2-\alpha)(d-2)}{R^4}\,(D\gamma_l\,-\,x_l)
\;,
$$
and
$$
\sum_{k=1}^d
[A^\alpha_k,F_{l,k}^\alpha]
\;=\;
\frac{4\alpha(\alpha^2-\alpha)(d-2)}{R^4}\,(D\gamma_l\,-\,x_l)
\;.
$$
In particular, for $d>2$ the Yang-Mills field equations \eqref{eq-YM} for $F^\alpha$ hold if and only if $\alpha\in\{0,\frac{1}{2},1\}$.
\end{proposi}

\noindent {\bf Proof.} 
Both identities follow from algebraic manipulations which require care and patience, but no creativity.
\hfill $\Box$

\subsection{Action of the special orthogonal group on the Clifford algebra}
\label{sec-Pin}

The special orthogonal group naturally acts on the Clifford algebra $\CM_d$ via
$
(O,\gamma_v)\;\mapsto\;\gamma_{Ov}
\;
$
and the multiplicative extension to products of $\gamma_v$'s and thus the whole Clifford algebra. This action is implemented by the restriction to the subgroup $\mbox{\rm SO}(d)$ of the group action of $ \mbox{\rm Pin}(d)$ on $\CM_d$ by adjunction \cite{LM}, as will be explained in some detail next. The crucial fact, following from a straightforward calculation, is that
$$
\gamma_v\,\gamma_w\,\gamma_v
\;=\;
\gamma_{R_vw}
\;,
$$
where $\|v\|=1$ and the reflection $R_v$ is defined by $R_{v}(\lambda v+w^\perp)= \lambda v-w^\perp$ for $\lambda\in\RM$ and $w^\perp\in\RM^d$ orthogonal to $v$. Note that $R_v$ is orthogonal and squares to the identity, and has $1$ as a simple eigenvalue. Hence $-R_v$ is a reflection in the conventional sense (leaving a hyperplane invariant). In order to further extend the action, let us recall next that the Clifford algebra is equipped with a linear anti-involution $i$ satisfying $i(\gamma_v)=\gamma_v$, $\;i^2=1$, $\;i(\gamma\gamma')=i(\gamma')i(\gamma)$. This corresponds simply to the transpose (in a faithful matrix representation of the Clifford algebra). Then, given a sequence $v_1,\ldots,v_k$ of unit vectors,
$$
(\gamma_{v_1}\cdots \gamma_{v_k})\,\gamma_w\,i(\gamma_{v_1}\cdots \gamma_{v_k})
\;=\;
\gamma_{R_{v_1}\cdots R_{v_k}w}
\;.
$$
On the other hand, every $O\in\mbox{\rm SO}(d)$ can be decomposed into an even number (more precisely, $d$ of them for even $d$ and $d-1$ for odd $d$) of reflections $O=R_{v_1}\cdots R_{v_k}$. Indeed, by the spectral theorem $O$ can be diagonalized into a block diagonal matrix with $2\times 2$ blocks given by rotations, and each such rotation block $r$ can be factorized into two reflections $r=\sigma_3(\sigma_3 r)$. In conclusion, for any $O\in\mbox{\rm SO}(d)$ one can set $g_O=\gamma_{v_1}\cdots\gamma_{v_k}$, which is a lift of $O$ into $ \mbox{\rm Spin}(d)$. Then the above reads
\begin{equation}
\label{eq-OGammaRep}
\gamma_{Ow}
\;=\;
g_O\,\gamma_w \,i(g_O)
\;.
\end{equation}
While all this is independent of the representation of $\CM_d$, we here only wrote it out in one given representation on $\CM^{2N}$, so that $O\in\mbox{\rm SO}(d)\mapsto g_O$ is a unitary  representation of $\mbox{\rm SO}(d)$ on $\CM^{2N}$.  

\vspace{.15cm}

The group $\mbox{\rm SO}(d)$ also naturally acts on $\RM^d$ and thus on $L^2(\RM^d)$ by
$$
(O\cdot \psi)(x)
\;=\;
\psi(O^*x)
\;,
\qquad
O\in \mbox{\rm SO}(d)\;,\;\;\psi\in L^2(\RM^d)
\;.
$$
Without the monopole, that is $\alpha=0$, one can drop the fibre $\CM^{2N}$.   Then one readily checks on the one parameter group that $Oe^{t\nabla^0_v}O^*=e^{t\nabla^0_{Ov}}$ and so,
\begin{equation}
\label{eq-ONabla0Rep}
O\,\nabla^0_v \,O^*\;=\;\nabla^0_{Ov}\;.
\end{equation}
Now let us extend the action  of $\mbox{\rm SO}(d)$ to $L^2(\RM^d,\CM^{2N})$ by tensoring with the unitary representation $O\in\mbox{\rm SO}(d)\mapsto g_O$. This action is denoted by $\hat{g}_O=O\otimes g_O$. As $A_v$ is a multiplication operator on $L^2(\RM^d,\CM^{2N})$, one has
\begin{equation}
\label{eq-SpinCov}
\hat{g}_OA^\alpha_v(x)\,\hat{g}_O^*
\;=\;
\hat{g}_O\frac{\alpha}{2}\;\frac{[\gamma_x,\gamma_v]}{R^2}\,\hat{g}_O^*
\;=\;
\frac{\alpha}{2}\;\frac{[g_O\gamma_{O^*x} g_O^*,g_O\gamma_vg_O^*]}{R^2}
\;=\;
\frac{\alpha}{2}\;\frac{[\gamma_x,\gamma_{Ov}]}{R^2}
\;=\;A^\alpha_{Ov}(x)
\;,
\end{equation}
or alternatively $A^\alpha_v(Ox)=g_OA^\alpha_{O^*v}(x)g_O^*$. One deduces upon combination with \eqref{eq-ONabla0Rep}
\begin{equation}
\label{eq-ONablaRep}
\hat{g}_O\nabla^\alpha_v \,\hat{g}_O^*\;=\;\nabla^\alpha_{Ov}\;.
\end{equation}
Also the field strength $\FieldStrength^\alpha_{v,w}(x)=[\nabla_v^\alpha,\nabla_w^\alpha]$ satisfies a similar covariance relation, namely 
\begin{equation}
\label{eq-OFieldRep}
\hat{g}_O\FieldStrength^\alpha_{v,w}\,\hat{g}_O^*
\;=\;
\FieldStrength^\alpha_{Ov,Ow}
\;,
\qquad
\FieldStrength^\alpha_{v,w}(Ox)
\;=\;
g_O\FieldStrength^\alpha_{O^*v,O^*w}(x)g_O^*
\;.
\end{equation}
%

\subsection{Total field strength}
\label{sec-TotalFieldStrength}

Let us introduce the differential $2n$-forms on $\RM^d\setminus\{0\}$
\begin{equation}
\label{eq-TotCurv}
\Omega_{2n}^\alpha
\;=\;
\Tr\big(G(\Ff^\alpha)^{\wedge n}\big)
\end{equation}
where $\Ff^\alpha$ is the curvature $2$-form associated to $\FieldStrength^\alpha$ by \eqref{eq-ConnForm}. For even $d$ so that one also disposes of $\Gamma$, let us set
\begin{equation}
\label{eq-TotCurv2}
\Omega_{2n+1}^\alpha
\;=\;
\Tr\big(\Gamma\Aa^\alpha\wedge (\Ff^\alpha)^{\wedge n}\big)
\;,
\end{equation}
here with a connection $1$-form $\Aa^\alpha$ defined by \eqref{eq-ConnForm}. These are $(2n+1)$-forms on $\RM^d\setminus\{0\}$. In this section, these forms are evaluated and then those of suitable degree are integrated over the unit sphere (or equivalently any homotopic surface). Towards the end of the section, it will be argued that these integrals are the total field strength. Let us begin by examining the behaviour of these forms under rotations and radial scaling. The equations \eqref{eq-ONablaRep} and \eqref{eq-OFieldRep} imply transformation laws for the connection $1$-form and the curvature $2$-form under the maps $O:\RM^d\setminus\{0\}\to\RM^d\setminus\{0\}$:
\begin{equation}
\label{eq-InvarForms}
O^*\,\Aa^\alpha\;=\;g_O\,\Aa^\alpha\,g_O^*
\;,
\qquad
O^*\,\Ff^\alpha\;=\;g_O\,\Ff^\alpha\,g_O^*
\;.
\end{equation}
Indeed, with matrix entries $O_{k,l}$ of $O$,
$$
(O^*\,\Aa^\alpha)(x)
\,=\,
\sum_{k,l=1}^d\,g_O\,A_{O^*e_k}(x)\,g_O^*\,O_{k,l}{\rm d}x_l
\,=\,
\sum_{k,l,m=1}^d\,g_O\,O_{k,m}\,A_{m}(x)\,g_O^*\,O_{k,l}{\rm d}x_l
\,=\,g_O\,\Aa^\alpha(x)\,g_O^*
\,,
$$
and similarly for the curvature form. From the invariance properties \eqref{eq-InvarForms} of the connection $1$-form and curvature $2$-form combined with the invariance of the trace, one deduces
\begin{equation}
\label{eq-CurvInvariance}
O^*\,\Omega_{2n}^\alpha
\;=\;
\Omega_{2n}^\alpha
\;,
\qquad
O^*\,\Omega_{2n+1}^\alpha
\;=\;
\Omega_{2n+1}^\alpha
\;.
\end{equation}
As to radial scaling $S_r:\RM^d\setminus\{0\}\to\RM^d\setminus\{0\}$ defined by $S_r(x)=rx$ for $r>0$, the equations \eqref{eq-AFormula3} and \eqref{eq-FieldStrengthScale} imply $S_r^*\,\Aa^\alpha=\Aa^\alpha$ and $S_r^*\,\Ff^\alpha\;=\;\Ff^\alpha$. As also $S_r^*G=G$, it follows that
\begin{equation}
\label{eq-CurvInvariance2}
S_r^*\,\Omega_{2n}^\alpha
\;=\;
\Omega_{2n}^\alpha
\;,
\qquad
S_r^*\,\Omega_{2n+1}^\alpha
\;=\;
\Omega_{2n+1}^\alpha
\;.
\end{equation}
It is hence sufficient to restrict the forms on the unit sphere $\SM^{d-1}$. For the form $\Omega^\alpha_{d-1}$ of maximal degree on $\SM^{d-1}$, the rotational invariance \eqref{eq-CurvInvariance} implies that the restriction is proportional to the volume form $\nu_{d-1}$ on $\SM^{d-1}$. In particular, the form  $\Omega^\alpha_{d-1}$ is closed on all $\RM^d\setminus\{0\}$. Let us evaluate the proportionality constant on $\SM^{d-1}$.

\begin{proposi}
\label{prop-CurvForm}
If the Clifford representation is irreducible, the restriction of $\Omega^\alpha_{d-1}$ to $\SM^{d-1}$ is given by
$$
\Omega^\alpha_{d-1}
\;=\;
\left\{
\begin{array}{cc}
\big(2\imath(\alpha-\alpha^2)\big)^{\frac{d-1}{2}}\,
(d-1)! \,\nu_{d-1}\;,
& d\;\mbox{ odd}\;,
\\
\\
-2\imath\alpha\big(2\imath(\alpha-\alpha^2)\big)^{\frac{d}{2}-1}\,
(d-1)! \,\nu_{d-1}\;,
& d\;\mbox{ even}\;.
\end{array}
\right.
$$
\end{proposi}

\noindent {\bf Proof.} Due to the rotation invariance of $\Omega^\alpha_{d-1}$, it is sufficient to evaluate $\Omega^\alpha_{d-1}(x)$ at one point $x\in\SM^{d-1}$ which we choose to be the unit vector $e_d$. Starting from \eqref{eq-MonopoleField} with $R=1$ and $x=e_d$, one first finds
$$
\Ff^\alpha(e_d)
\;=\;
(\alpha^2-\alpha)\sum_{k,l=1}^d
\big(\gamma_l\gamma_k
\;+\;
\gamma_d(\delta_{k,d}\gamma_l-\delta_{l,d}\gamma_k)\big)
{\rm d}x_k\wedge{\rm d}x_l
\;=\;
(\alpha-\alpha^2)\sum_{k,l=1}^{d-1}
\gamma_k\gamma_l\,
{\rm d}x_k\wedge{\rm d}x_l
\;.
$$
Therefore for odd $d$ and thus $n=\frac{d-1}{2}$, one has $G(e_d)=\gamma_d$ and thus
\begin{align*}
\Omega_{d-1}^\alpha(e_d)
&
\;=\;
(\alpha-\alpha^2)^{\frac{d-1}{2}}\,\sum_{k_1,\ldots,k_{d-1}=1}^{d-1}
\Tr\big(\gamma_d
\gamma_{k_1}\cdots \gamma_{k_{d-1}}\big)
\,
{\rm d}x_{k_1}\wedge\ldots\wedge{\rm d}x_{k_{d-1}}
\\
&
\;=\;
(\alpha-\alpha^2)^{\frac{d-1}{2}}\,
(d-1)!\,\Tr(\gamma_1\cdots\gamma_{d})\,
{\rm d}x_{1}\wedge\ldots\wedge{\rm d}x_{d-1}
\;.
\end{align*}
Now due to our choice of the Clifford representation, $\Tr(\gamma_1\cdots\gamma_{d})=  \imath^{\frac{d-1}{2}}\,\Tr(\one)=(2\imath)^{\frac{d-1}{2}}$ and this implies the claim for odd $d$. For even $d$, let us note that
$$
\Aa^\alpha(e_d)
\;=\;
\alpha \sum_{k=1}^{d-1}\gamma_d\gamma_k\,{\rm d}x_k
\;.
$$
Hence
\begin{align*}
\Omega_{d-1}^\alpha(e_d)
&
\;=\;
\alpha(\alpha-\alpha^2)^{\frac{d}{2}-1}\,\sum_{k=1}^{d-1}\sum_{k_1,\ldots,k_{d-2}=1}^{d-1}
\Tr\big(\Gamma\,\gamma_d\gamma_k
\gamma_{k_1}\cdots \gamma_{k_{d-2}}\big)
\,
{\rm d}x_{k}\wedge{\rm d}x_{k_1}\wedge\ldots\wedge{\rm d}x_{k_{d-2}}
\\
&
\;=\;
\alpha(\alpha-\alpha^2)^{\frac{d}{2}-1}\,
(d-1)!\,\Tr(\Gamma\gamma_d\gamma_1\cdots\gamma_{d-1})\,
{\rm d}x_{1}\wedge\ldots\wedge{\rm d}x_{d-1}
\;.
\end{align*}
Finally $\Tr(\Gamma\gamma_d\gamma_1\cdots\gamma_{d-1})=-\Tr(\Gamma\gamma_1\cdots\gamma_{d})=-(2\imath)^{\frac{d}{2}}$.
\hfill $\Box$

\vspace{.15cm}

Next recall that the double factorial for odd $d$ is given by 
\begin{equation}
\label{eq-DoubleFact}
d\,!!\;=\;d(d-2)\cdots 3\cdot 1\;=\;\frac{d\,!}{\frac{d-1}{2}!\, 2^{\frac{d-1}{2}}}
\;=\;\frac{(d+1)!}{\frac{d+1}{2}!\, 2^{\frac{d+1}{2}}}
\;.
\end{equation}
%

\begin{coro}
\label{coro-CurvForm}
If the Clifford representation is irreducible, one has
$$
\int_{\SM^{d-1}}\Omega^\alpha_{d-1}
\;=\;
\left\{
\begin{array}{cc}
2\big(8\imath(\alpha-\alpha^2)\pi\big)^{\frac{d-1}{2}}\,
\tfrac{d-1}{2}!
\;,
& d\;\mbox{ odd}\;,
\\
\\
-4\imath\alpha\pi\big(4\imath(\alpha-\alpha^2)\pi\big)^{\frac{d}{2}-1}\,(d-1)!!\;,
& d\;\mbox{ even}\;.
\end{array}
\right.
$$
\end{coro}

\noindent {\bf Proof.}  In view of Proposition~\ref{prop-CurvForm}, one needs to use the expression for the volume of $\SM^{d-1}$:
$$
\nu_{d-1}(\SM^{d-1})
\;=\;
\frac{2\,\pi^{\frac{d}{2}}}{\Gamma(\frac{d}{2})}
\;=\;
\left\{
\begin{array}{cc}
\frac{2\,(2\pi)^{\frac{d-1}{2}}}{(d-2)!!}
\;,
& d\;\mbox{ odd}\;,
\\
\\
\frac{2\,\pi^{\frac{d}{2}}}{(\frac{d}{2}-1)!}
\;,
& d\;\mbox{ even}\;.
\end{array}
\right.
$$
Careful evaluation using the identity \eqref{eq-DoubleFact} shows the claim. \hfill $\Box$

\vspace{.15cm}

Because the form $\Omega^\alpha_{d-1}$ is closed in $\RM^d\setminus\{0\}$, Stokes' theorem shows that the integral in Corollary~\ref{coro-CurvForm} is equal to $\int_{\Mm_{d-1}}\Omega^\alpha_{d-1}$ for any closed $(d-1)$-dimensional manifold $\Mm_{d-1}$ which is homotopic to $\SM^{d-1}$ in $\RM^d\setminus\{0\}$. Finally, let us comment on why $\int_{\SM^{d-1}}\Omega^\alpha_{d-1}$ should be interpreted as the total field strength of the monopole. In odd dimension, one may be tempted to consider $\Tr\big((\Ff^\alpha)^{\wedge n}\big)$ instead of \eqref{eq-TotCurv}. However, the proof of Proposition~\ref{prop-CurvForm} shows that $\Tr\big((\Ff^\alpha)^{\wedge n}\big)=0$. The motivation for considering $\Omega^\alpha_{d-1}$ comes from the next subsection where it is shown that the forms at $\alpha=\frac{1}{2}$ are the integrands in the higher Chern and winding numbers. They are thus of geometrical origin. The non-abelian nature of the fields makes it difficult to inteprete $\Omega^\alpha_{d-1}(x)$ as the field pointing outwards at $x\in\SM^{d-1}$ multiplied by the surface element, but one can interprete $G(x)$ as a vector pointing outwards with non-commutative entries given by Clifford matrices and the trace over these matrix degrees of freedom as a suitable state. In $d=3$ and for $x=e_3$, one has $F_{1,2}(e_3)=2\imath (\alpha-\alpha^2)\gamma_3$ and $F_{2,3}(e_3)=F_{1,3}(e_3)=0$ (see Subsection~\ref{sec-Monopole234D}) which, in (abelian) electromagnetism, are (twice) the magnetic fields $B_3$, $B_1$ and $B_2$ respectively. Extracting the component pointing outwards is precisely given by the state $\gamma\mapsto \Tr(G(e_3)\gamma)=\Tr(\gamma_3\gamma)$. 

\subsection{Half-flux monopole and its charge}
\label{sec-HalfFlux}

Proposition~\ref{prop-YM} already showed that half-flux, $\alpha=\frac{1}{2}$, is a special value. Here it is shown next that the half-flux field strength is indeed of geometric origin, namely it is the connection of the vector bundle given by the positive spectral projection $\Hardy $ of the Dirac operator. More precisely, let us consider the projections
\begin{equation}
\Hardy (x)\;=\;\frac{1}{2}\big(\one\,+\,G(x)\big)
\;,
\qquad
x\in\RM^d\setminus\{0\}
\;.
\label{eq-DiracSpecProj}
\end{equation}
As for any vector bundle specified by a projection, the connection is $[\Hardy (x),\partial \Hardy (x)]$. Since $G(x)=|x|^{-1}\langle x|\gamma\rangle$, one readily checks by comparing with \eqref{eq-AFormula} that
$$
A^{\frac{1}{2}}(x)
\;=\;
[\Hardy (x),\partial \Hardy (x)]
\;.
$$
The vector bundle is radially symmetric:
$
\Hardy (rx)\;=\;\Hardy (x)
\;,
\  r>0\;.
$
It is hence fixed by its values on the unit sphere $\SM^{d-1}$. Let us furthermore note that for even $d$, the projection (and the associated vector bundle) is chiral in the sense that
$
\Gamma\,\Hardy (x)\,\Gamma\;=\;\one\,-\,\Hardy (x)
\;.
$
This means that $\Hardy (x)$ is of the form
\begin{equation}
\label{eq-ChiralPMatRep}
\Hardy (x)\;=\;\frac{1}{2}
\begin{pmatrix}
\one & W(x)^* \\ W(x) & \one
\end{pmatrix}
\;,
\end{equation}
wherethe $2\times 2$ matrix is in the spectral representation of $\Gamma=\diag(\one,-\one)$ and $W(x)$ is a unitary matrix. See also \cite{DG} for more information on chiral vector bundles.

\vspace{.15cm}

As the half-flux stems from a vector bundle described by the projection $\Hardy $, also the connection $1$-form and curvature $2$-form can be expressed in terms of $\Hardy $:
$$
\Aa^{\frac{1}{2}}\;=\;[\Hardy ,{\rm d}\Hardy ]\;,
\qquad
\Ff^{\frac{1}{2}}
\;=\;{\rm d} \Hardy \wedge {\rm d} \Hardy 
\;.
$$
Hence, using $\Hardy {\rm d}\Hardy ={\rm d}\Hardy (\one-\Hardy )$ and $(\one-\Hardy ){\rm d}\Hardy ={\rm d}\Hardy \,\Hardy $, the differential forms defined in \eqref{eq-TotCurv} are
\begin{equation}
\label{eq-Omega2n}
\Omega_{2n}^{\frac{1}{2}}
\;=\;
\Tr\big((\Hardy -(\one-\Hardy ))({\rm d}\Hardy \wedge {\rm d}\Hardy )^{\wedge n}\big)
\;=\;
2\,\Tr\big(\Hardy ({\rm d}\Hardy \wedge {\rm d}\Hardy )^{\wedge n}\big)
\;.
\end{equation}
For even $d$, one gets for \eqref{eq-TotCurv2} that
\begin{equation}
\label{eq-Omega2n+1}
\Omega_{2n+1}^{\frac{1}{2}}
\;=\;
\Tr\big(\Gamma(\Hardy {\rm d}\Hardy -{\rm d}\Hardy \,\Hardy )({\rm d}\Hardy \wedge {\rm d}\Hardy )^{\wedge n}\big)
\;=\;
2\,\Tr\big(\Gamma \Hardy ({\rm d}\Hardy )^{\wedge 2n+1}\big)
\;,
\end{equation}
because ${\rm d}\Hardy \,\Gamma=-\Gamma {\rm d}\Hardy $ so that $\Tr(\Gamma ({\rm d}\Hardy )^{\wedge 2n+1})=0$. These forms will now be integrated over the sphere $\SM^{d-1}$. With suitable normalization, these integrals are the even or odd Chern numbers of the vector bundle  on the sphere $\SM^{d-1}$ specified by $\Hardy $.

\vspace{.15cm}

Let us recall the definition of these Chern numbers in a slightly more general context ({\it e.g.} \cite{GiS,Nak,PS2}), albeit with the dimension $d$  shifted by $1$ to fit our present context. Let $\Mm_{d-1}$ be a $(d-1)$-dimensional closed and compact Riemannian manifold and $x\in\Mm_{d-1}\mapsto \Hardy (x)$ be a differentiable map with values in the projections on $\CM^{2N}$ for some $N$. For even $d$, suppose also given a linear involution $\Gamma$ on  $\CM^{2N}$ such that $\Gamma \Hardy (x)\Gamma=\one-\Hardy (x)$, namely $\Hardy $ is chiral w.r.t. $\Gamma$.  Now the Chern numbers are
%
%
$$
\Ch_{d-1}(\Hardy )
\;=\;
\left\{
\begin{array}{cc}
\big(\tfrac{1}{2\pi\imath}\big)^{\frac{d-1}{2}}\,\tfrac{1}{\frac{d-1}{2}!}\int_{\Mm_{d-1}}\!\Tr\big(\Hardy ({\rm d}\Hardy )^{\wedge (d-1)}\big)\;,
& d\;\mbox{ odd}\;,
\\
\\
\big(\tfrac{1}{\pi\imath}\big)^{\frac{d}{2}}\,\tfrac{-1}{(d-1)!!}\int_{\Mm_{d-1}}\!\Tr\big(\Gamma \Hardy ({\rm d}\Hardy )^{\wedge (d-1)}\big)\;,
& d\;\mbox{ even}\;.
\end{array}
\right.
$$
For odd $d$, our notation deviates from the standard one \cite{GiS,Nak} in which one denotes $\Ch_{d-1}(\Hardy )$ rather by $\Ch_{\frac{d-1}{2}}(\Hardy )$, and on top of this notational shift the normalization constant differs from the usual one ({\it e.g.} \cite{GiS,Nak,PS2}) by a sign $(-1)^{\frac{d-1}{2}}$. With the present choice, the Chern number of $\Hardy$ is equal to $1$, see Proposition~\ref{prop-charge} below. For even $d$, the odd Chern numbers are also called higher winding numbers. To realize why and to show that the normalization constant is chosen exactly as in \cite{PS,PS2}, let us go into the spectral representation of $\Gamma$ so that \eqref{eq-ChiralPMatRep} holds.  As ${\rm d}W^*=-W^*\,{\rm d}W\,W^*$,
\begin{align*}
\Tr\big(\Gamma \Hardy ({\rm d}\Hardy )^{\wedge (d-1)}\big)
& 
\;=\;
\frac{1}{2^d}\,
\Tr
\left(
\begin{pmatrix}
\one & 0 \\ 0 & -\one
\end{pmatrix}
\begin{pmatrix}
\one & W^* \\ W & \one
\end{pmatrix}
\begin{pmatrix}
0 & {\rm d}W^* \\ {\rm d} W & 0
\end{pmatrix}^{\wedge (d-1)}
\right)
\\
&
\;=\;
\frac{1}{2^{d-1}}\,(-1)^{\frac{d-2}{2}}\,\Tr\big((W^*{\rm d}W)^{\wedge (d-1)}\big)
\;,
\end{align*}
so that 
$$
\Ch_{d-1}(\Hardy )
\;=\;
\big(\tfrac{\imath}{\pi}\big)^{\frac{d}{2}}\,\tfrac{1}{2^{d-1}(d-1)!!}\int_{\Mm_{d-1}}\!
\Tr\big((W^*{\rm d}W)^{\wedge (d-1)}\big)\;,
\qquad 
d\;\mbox{ even}\;.
$$
For this reason, one can also write $\Ch_{d-1}(W)$ instead of $\Ch_{d-1}(\Hardy )$ (as in \cite{PS,PS2}).

\vspace{.15cm}

These definitions can now be applied to the projections \eqref{eq-DiracSpecProj} restricted to the manifold $\SM^{d-1}$. Note that for odd $d-1$, namely even $d$, the projection is indeed chiral. Hence $\Ch_{d-1}(\Hardy )$ is defined for all dimensions $d$. This number is called the (topological) charge of the half-flux monopole. For odd $d$, the following also follows from the results of \cite{GiS}.

\begin{proposi}
\label{prop-charge}
For an irreducible Clifford representation the half-flux monopole has charge $1$.
\end{proposi}

\noindent {\bf Proof.} 
For odd $d$, using \eqref{eq-Omega2n} and Corollary~\ref{coro-CurvForm} one finds
$$
\Ch_{d-1}(\Hardy )
\;=\;
\big(\tfrac{1}{2\pi\imath}\big)^{\frac{d-1}{2}}\,\tfrac{1}{\frac{d-1}{2}!}
\int_{\SM^{d-1}}\!
\tfrac{1}{2}\,\Omega^{\frac{1}{2}}_{d-1}
\;=\;
\big(\tfrac{1}{2\pi\imath}\big)^{\frac{d-1}{2}}\,\tfrac{1}{\frac{d-1}{2}!}
\;
\big(2\imath\pi\big)^{\frac{d-1}{2}}\,
\tfrac{d-1}{2}!
\;=\;
1
\;.
$$
In the even dimensional case, 
$$
\Ch_{d-1}(\Hardy )
\;=\;
\big(\tfrac{1}{\imath\,\pi}\big)^{\frac{d}{2}}\,\tfrac{-1}{(d-1)!!}
\int_{\SM^{d-1}}\!
\tfrac{1}{2}\,\Omega^{\frac{1}{2}}_{d-1}
\;=\;
\big(\tfrac{1}{\imath\,\pi}\big)^{\frac{d}{2}}\,\tfrac{-1}{(d-1)!!}
\;
(-\imath\pi)\big(\imath\pi\big)^{\frac{d}{2}-1}\,(d-1)!!
\;=\;
1
\;,
$$
concluding the proof.
\hfill $\Box$

\subsection{Examples of monopoles in dimension $d=2,3,4$}
\label{sec-Monopole234D}

Let us begin with the case $d=2$. Then $\gamma_1=\sigma_1$, $\gamma_2=\sigma_2$ and $\Gamma=\sigma_3$ are expressed in terms of the Pauli matrices. The Dirac operator is
$$
D
\;=\;
X_1\sigma_1+X_2\sigma_2
\;=\;
\begin{pmatrix}
0 & X_1\,-\,\imath\,X_2 \\ X_1\,+\,\imath\,X_2 & 0
\end{pmatrix}
\;,
$$
and the gauge potential is
$$
A_1^\alpha(x)
\;=\; 
\frac{\imath\,\alpha}{R^2}
\begin{pmatrix}
-\,x_2 & 0 \\ 0 & x_2
\end{pmatrix}
\;,
\qquad
A_2^\alpha(x)
\;=\; 
\frac{\imath\,\alpha}{R^2}
\begin{pmatrix}
x_1 & 0 \\ 0 & -\,x_1
\end{pmatrix}
\;,
$$
where $R^2=(x_1)^2+(x_2)^2$. One checks that the field strength vanishes away from the origin, that is $
\FieldStrength^\alpha_{1,2}(x)=0$ for $x\not=0$. Now integration of \eqref{eq-NDGL} leads to 
$$
N_1(x)
\;=\;
\begin{pmatrix}
e^{\imath\,\alpha\arctan(\frac{x_1}{x_2})} & 0 \\ 0 & e^{-\imath\,\alpha\arctan(\frac{x_1}{x_2})}
\end{pmatrix}
\;,
\qquad
N_2(x)
\;=\;
\begin{pmatrix}
e^{-\imath\,\alpha\arctan(\frac{x_2}{x_1})} & 0 \\ 0 & e^{\imath\,\alpha\arctan(\frac{x_2}{x_1})}
\end{pmatrix}
\;.
$$
From this, $M^\alpha_1(x)$ and $M^\alpha_2(x)$ can readily be deduced. The first component of these formulas and all the associated claims of Proposition~\ref{prop-MonTrans} coincide (up to a sign change) with those in \cite{DS} which were deduced there by more complicated algebraic manipulations.  Let us note that the second component is obtained by replacing $\alpha$ by $-\alpha$. This doubling will be further discussed in Section~\ref{sec-SFHam}.

\vspace{.15cm}

Next let us move on to $d=3$. Then $\gamma_1=\sigma_1$, $\gamma_2=\sigma_2$ and $\gamma_3=\sigma_3$ and the gauge potential is given by
$$
A_1^\alpha(x)
\;=\; 
\frac{\imath\alpha(x_3\sigma_2-x_2\sigma_3)}{R^2}
\;,
\qquad
A_2^\alpha(x)
\;=\; 
\frac{\imath\alpha(x_1\sigma_3-x_3\sigma_1)}{R^2}
\;,
\qquad
A_3^\alpha(x)
\;=\; 
\frac{\imath\alpha(x_2\sigma_1-x_1\sigma_2)}{R^2}
\;,
$$
where $R^2=(x_1)^2+(x_2)^2+(x_3)^2$. Careful evaluation of \eqref{eq-MonopoleField} gives the non-abelian magnetic field of the Wu-Yang monopole: 
$$
\FieldStrength^\alpha_{1,2}
\;=\;
\frac{2(\alpha^2-\alpha)x_3D}{\imath\,R^4}
\;,
\qquad
\FieldStrength^\alpha_{2,3}
\;=\;
\frac{2(\alpha^2-\alpha)x_1D}{\imath\,R^4}
\;,
\qquad
\FieldStrength^\alpha_{3,1}
\;=\;
\frac{2(\alpha^2-\alpha)x_2D}{\imath\,R^4}
\;.
$$
For $\alpha=\frac{1}{2}$, this corresponds to the Wu-Yang SU$(2)$-monopole \cite{WY} (for parameter $\Phi=-1$ in the notation of \cite{WY}, see also equation (5.4) in \cite{Shn}).  As explained in Section 5.1.1. and Appendix~C in \cite{Shn}, this Wu-Yang monopole is also a non-abelian gauge for the Dirac monopole, which has an abelian, but singular gauge.

\vspace{.15cm}

Finally in dimension $d=4$ the spinor space is $\CM^4=\CM^2\otimes\CM^2$. As a representation of the Clifford algebra we choose
$$
\gamma_1\,=\,\sigma_1\otimes\sigma_2
\;,
\quad
\gamma_2\,=\,\sigma_2\otimes\sigma_2
\;,
\quad
\gamma_3\,=\,\sigma_3\otimes\sigma_2
\;,
\quad
\gamma_4\,=\,\sigma_0\otimes\sigma_1
\;.
$$
The chiral symmetry is $\Gamma=\sigma_0\otimes\sigma_3$. Here $\sigma_0$ is the identity. The Dirac operator is chiral and thus off-diagonal in the grading of $\Gamma$, and $A^\alpha_k$ and $\FieldStrength^\alpha_{l,k}$ are block-diagonal and their block diagonal terms can be written in terms of the Pauli matrices again. It is straightforward, albeit tedious, to write out the non-abelian vector potential and field strength. One finds, for example,
$$
A_1^\alpha
\;=\;
\frac{\imath\alpha}{R^2}
\begin{pmatrix}
x_4\sigma_1+x_3\sigma_2- x_2\sigma_3 & 0 
\\
0 & -x_4\sigma_1+x_3\sigma_2- x_2\sigma_3 
\end{pmatrix}
\;,
$$
and
$$
A_2^\alpha
\;=\;
\frac{\alpha}{R^2}
\begin{pmatrix}
x_1\sigma_3-x_3\sigma_1+ x_4\sigma_2 & 0 
\\
0 & x_1\sigma_3-x_3\sigma_1+x_4\sigma_2 
\end{pmatrix}
\;,
$$
so that 
$$
\FieldStrength^\alpha_{1,2}
\;=\;
\frac{2(\alpha^2-\alpha)}{R^4}
\begin{pmatrix}
(x_1x_3+x_2x_4)\sigma_1+(x_2x_3-x_1x_4)\sigma_2+(x_3^2+x_4^2)\sigma_3 \hspace{3cm} 0 \;\; 
\\
\;\;0 \hspace{3cm} (x_1x_3-x_2x_4)\sigma_1+(-x_2x_3+x_1x_4)\sigma_2+(x_3^2+x_4^2)\sigma_3 
\end{pmatrix}
\;.
$$

For $d=5$, we suspect that the SU$(5)$-monopole \eqref{eq-MonopoleDef} with $\alpha=\frac{1}{2}$ is gauge equivalent to the monopole constructed by Yang \cite{Yan}. A similar statement may also hold for the monopole in dimension $d=9$ presented in \cite{LN}.

\section{Non-abelian monopole translations}
\label{sec-MonopolTrans}

The aim of this section is to construct monopole translations on a lattice Hilbert space. These will then lead to a one-parameter family of Hamiltonians indexed by the flux $\alpha$ for which the spectral flow can be studied. We follow the strategy of Arai \cite{Ara} for the construction of magnetic translations, namely to first build these translations in the continuum and then descend them to the lattice. A crucial difference w.r.t. \cite{Ara} is, however, that the gauge fields and thus also the Peierls factors are non-abelian here. In the present work, magnetic fields from an abelian $\mbox{\rm U}(1)$-gauge potential are not taken into account for the  sake of simplicity, but this does not alter the constructions in any essential manner.

\subsection{Selfadjointness of covariant derivatives}
\label{sec-Selfadjointness}

Here the covariant derivatives $\nabla^\alpha_v$ introduced in Section~\ref{sec-CovDeriv} are studied from a more functional analytic point of view. For that purpose let us consider the line $L_{x,v}=\{x+tv\,:\,t\in\RM\}$ through $x$ in the direction $v$ where $x\in\RM^d$ is supposed to satisfy $\langle x|v\rangle=0$. For $x\not=0$, the map $x'\in  L_{x,v}\mapsto A^\alpha_v(x')$ is smooth and takes values in the skew-adjoint matrices. Now let us consider the dense subspace $C^\infty_0(\RM^d\setminus L_{0,v},\CM^{2N})\subset L^2(\RM^d,\CM^{2N})$ of smooth functions vanishing on the line $L_{0,v}$ through the origin. One readily checks that $\imath\nabla^\alpha_v$ is a symmetric operator on $C^\infty_0(\RM^d\setminus L_{0,v},\CM^{2N})$.

\begin{proposi}
\label{prop-SelfAdjoint}
For $v\in\RM^d$ and $\alpha\in\RM$, the operators $\imath\nabla^\alpha_v$ have unique selfadjoint extensions. 
\end{proposi}

\noindent {\bf Proof.} For $x\not=0$ satisfying $\langle x|v\rangle=0$, let us first solve the following first order ordinary differential equation on $L_{x,v}$:
\begin{equation}
\label{eq-NDGL}
\partial_v\,N_v^\alpha(x')\;=\;-\,A^\alpha_v(x')\,N_v^\alpha(x') 
\;,\qquad
N_v^\alpha(x)\;=\;\one_{2N}
\;,
\end{equation}
or equivalently
$$
\partial_tN_v^\alpha(x+tv)\;=\;-\,A^\alpha_v(x+tv)\,N_v^\alpha(x+tv) 
\;,\qquad
N_v^\alpha(x)\;=\;\one_{2N}
\;.
$$
As $A^\alpha_v(x')$ is skew-adjoint, the solution $t\in\RM\mapsto N^\alpha_v(x+tv)$ is a path of unitaries. As the coefficients decrease at infinity due to the bound \eqref{eq-ABound}, the solution converges to a fixed unitary matrix at $t=\pm \infty$.  Also $N_v^\alpha=N_{sv}^\alpha$ for all $s\not=0$ because $A^\alpha_{sv}=sA^\alpha_v$. Next let us introduce $M_v^\alpha(x',t)$ by
\begin{equation}
\label{eq-MDef}
M_v^\alpha(x',t)
\;=\;
N_v^\alpha(x')\,
N_v^\alpha(x'+tv)^*
\;.
\end{equation}
For $\psi\in C^\infty_0(\RM^d\setminus L_{0,v},\CM^{2N})$ let us then set
$$
\big(\MonopolTrans^\alpha_v(t)\psi\big)(x')
\;=\;
M_v^\alpha(x',t)\,\psi(x'+tv)
\;.
$$
Note that $\MonopolTrans^\alpha_v(t)$ sends  $C^\infty_0(\RM^d\setminus L_{0,v},\CM^{2N})$ into itself and  $t\mapsto \MonopolTrans^\alpha_v(t)\psi$ is continuous w.r.t. to the norm on $L^2(\RM^d,\CM^{2N})$. Moreover, one readily checks
$$
\big(\MonopolTrans^\alpha_v(s) (\MonopolTrans^\alpha_v(t)\psi)\big)(x')
\;=\;\big(\MonopolTrans^\alpha_v(t+s)\psi)\big)(x')
\;.
$$
Hence $t\in\RM\mapsto \MonopolTrans^\alpha_v(t)$ is a one-parameter family of densely defined isometries on $L^2(\RM^d,\CM^{2N})$. By continuity, they extend to a one-parameter family of unitaries on $L^2(\RM^d,\CM^{2N})$. Furthermore, a standard $3\epsilon$-argument shows that this family is strongly continuous. By Stone's theorem, the associated generator is selfadjoint (on its naturally associated maximal domain). A short calculation shows that indeed the generator is given by \eqref{eq-CovDer}:
$
\imath\nabla^\alpha_v
\;=\;
\imath\,\partial_t\MonopolTrans^\alpha_v(t)|_{t=0}
\;,
$
Similarly,
$$
N_v^\alpha\, \nabla^0_v \,(N_v^\alpha)^*
\;=\;
\nabla^\alpha_v
\;,
\qquad
e^{t\nabla^\alpha_v}
\;=\;
N_v^\alpha\, e^{t\nabla^0_v} \,(N_v^\alpha)^*
\;=\;
M_v^\alpha(\,.\,,t)\, e^{t\nabla^0_v}
\;,
$$
if one views $N_v^\alpha$ and $M_v^\alpha(\,.\,,t)$ as a unitary multiplication operators on $L^2(\RM^d,\CM^{2N})$.
\hfill $\Box$

\vspace{.15cm}

\subsection{Basic properties of non-abelian monopole translations}
\label{sec-MonopolTransDef}

As $\imath\nabla^\alpha_v$ is selfadjoint by Proposition~\ref{prop-SelfAdjoint}, it can be exponentiated to a one-parameter family 
\begin{equation}
\label{eq-NablaGroup}
t\in\RM\,\mapsto\,e^{t \nabla^\alpha_v}\;=\;e^{\nabla^\alpha_{tv}}
\end{equation}
of unitary operators on $L^2(\RM^d,\CM^{2N})$. Proposition~\ref{prop-MonTrans} below shows that these operators are translations, modified by a local matrix multiplication. This is similar to the magnetic translations where the multiplication is merely by an abelian phase factor (see {\it e.g.} \cite{Ara}), and therefore we also call $e^{t \nabla^\alpha_v}$ the non-abelian monopole translations. Moreover, the following proposition shows that the non-abelian monopole translations in different directions are connected by a unitary transformation which invokes the Spin-representation of the special orthogonal group $\mbox{\rm SO}(d)$ described in Subsection~\ref{sec-Pin}. 

\begin{proposi}
\label{prop-MonTrans}
Let $v\in\RM^d$ and $O\in \mbox{\rm SO}(d)$. The non-abelian monopole translations are of the form
\begin{equation}
\label{eq-AnsatzMagTrans}
(e^{\nabla^\alpha_v}\psi)(x)
\;=\;
M_v^\alpha(x)\,\psi(x+v)\;,
\qquad
\psi\in L^2(\RM^d,\CM^{2N})
\;,
\end{equation}
where $x\in \RM^d\setminus\{tv:t\in[-1,0]\}\mapsto M_v^\alpha(x)\in\mbox{\rm U}(2N)$ is continuous. This function cannot be extended continuously to the line segment $\{tv:t\in[-1,0]\}$, but values on this set of zero measure do not alter $M_v^\alpha$ as a unitary multiplication operator on $L^2(\RM^d,\CM^{2N})$. The function has the following normalization at infinity:
\begin{equation}
\label{eq-MagTransAsymp}
\lim_{|x|\to\infty} M_v^\alpha(x)\;=\;\one_{2N}
\;.
\end{equation}
The non-abelian phase factors satisfy the covariance property
\begin{equation}
\label{eq-MonpoleTransCov}
g_O\,M_v^\alpha(O^*x)\,g_O^*
\;=\;
M_{Ov}^\alpha(x)
\;.
\end{equation}
Moreover,
\begin{equation}
\label{eq-MagTransConj}
\DiracPhase \,e^{ \nabla_v^\alpha}\,\DiracPhase 
\;=\;e^{\nabla_v^{1-\alpha}}
\;.
\end{equation}
For even $d$ and the chirality operator $\Gamma$ of $D$, one also has
\begin{equation}
\label{eq-MonoTransGrad}
\Gamma\, M_v^\alpha(x)\,\Gamma
\;=\;
M_v^\alpha(x)
\;,
\qquad
\Gamma\,e^{\nabla_v^\alpha}\,\Gamma
\;=\;
e^{\nabla_v^\alpha}
\;.
\end{equation}
\end{proposi}

\noindent {\bf Proof.} We will use the notations and formulas of the proof of Proposition~\ref{prop-SelfAdjoint}, in particular, the solution $t\in\RM\mapsto N_v^\alpha(x+tv)$ of \eqref{eq-NDGL} for $x\not=0$ satisfying $\langle x|v\rangle=0$. These solutions are clearly continuous. Therefore also the functions
\begin{equation}
\label{eq-MDef2}
x\in\RM^d\setminus L_{0,v}
\;\mapsto\;
M_v^\alpha(x)
\;=\;
N_v^\alpha(x)\,
N_v^\alpha(x+v)^*
\;,
\end{equation}
are continuous. It was also already shown in the proof of Proposition~\ref{prop-SelfAdjoint} that  $e^{ \nabla^\alpha_v}=\MonopolTrans^\alpha_v(1)$ so that \eqref{eq-AnsatzMagTrans} holds. Furthermore, let us show that $M_v^\alpha(x)$ can be extended continuously to $L_{0,v}\setminus \{tv:t\in[-1,0]\}=\{tv:t\in\RM\setminus[-1,0]\}$. Indeed, for $t\in\RM\setminus[-1,0]$, one has
$$
\lim_{x\to tv}\,M_v^\alpha(x)\;=\;\one_{2N}
\;,
$$
because by the mean value theorem and \eqref{eq-NDGL}
$$
\|M_v^\alpha(x)\,-\,\one_{2N}\|
\;=\;
\|N_v^\alpha(x)\,-\,N_v^\alpha(x+v)\|
\;\leq\;
\sup_{y\in[x,x+v]}
\|\alpha\,A_v(y) N_v^\alpha(y)\|
\;=\;
|\alpha|\sup_{y\in[x,x+v]}
\|A_v(y)\|
\;,
$$
which using formula \eqref{eq-AFormula2} can be shown to vanish in the limit $x\to tv$ for $t\in\RM\setminus[-1,0]$. Also the statement on the discontinuity is elementary to check. As already pointed out in the proof of Proposition~\ref{prop-SelfAdjoint}, the functions $t\in\RM\mapsto N_v^\alpha(x+tv)$ converge to a fixed unitary matrix at $t=\pm \infty$. Thus by \eqref{eq-MDef2}
$$
\lim_{t\to\pm\infty}
M_v^\alpha(x+tv)
\;=\;
\one_{2N}
\;,
$$
which is a first instance of \eqref{eq-MagTransAsymp}. Substituting \eqref{eq-ONablaRep} into \eqref{eq-AnsatzMagTrans} directly leads to the covariance relation \eqref{eq-MonpoleTransCov}. Furthermore, \eqref{eq-MagTransConj} follows from \eqref{eq-NablaConj}, and the last claim from \eqref{eq-DiracChir}. Based on \eqref{eq-MonpoleTransCov}, one also deduces the general form of \eqref{eq-MagTransAsymp}.
\hfill $\Box$

\subsection{Non-abelian monopole translations on the lattice}
\label{sec-MonopolTransLatt}

In view of \eqref{eq-AnsatzMagTrans}, the operator $e^{\nabla^\alpha_{e_k}}$ is the translation by $e_k$ followed by a unitary multiplication operator. In particular, $e^{\nabla^\alpha_{e_k}}$ therefore leaves the lattice Hilbert space $\ell^2(\ZM^d,\CM^{2N})$ invariant. Hence the monopole translation $S^\alpha_k$ in direction $k$ on the lattice is morally the restriction of $e^{\nabla^\alpha_{e_k}}$ to the lattice Hilbert space. More formally, for $w\in\CM^{2N}$ specifying a localized state $w\otimes|n\rangle$ at site $n$,
\begin{equation}
\label{eq-SalphaDef}
S^\alpha_k\; w\otimes |n\rangle\;=\;M^\alpha_{e_k}(n-e_k)w\otimes |n-e_k\rangle
\;.
\end{equation}
Due to Proposition~\ref{prop-MonTrans}, the unitary $M^\alpha_{e_k}(n-e_k)$ is well-defined as the value of a continuous function on all sites except at $n=e_k$ and $n=0$. At these points, particular unitary matrices have to be chosen. Our criterion is that \eqref{eq-MagTransConj} remains valid for the magnetic translations, namely that \eqref{eq-MonopoleTransConj} below holds. This requires us also to define a selfadjoint unitary matrix $\DiracPhase(0)$ because again \eqref{eq-DGdef} does not provide this matrix. Then $\DiracPhase $ is a selfadjoint unitary multiplication operator on $\ell^2(\ZM^d,\CM^{2N})$.

\begin{proposi}
\label{prop-SalphaDef}
One can choose a selfadjoint unitary matrix $\DiracPhase(0)$ and unitary matrices $M^\alpha_{e_k}(0)$ and $M^\alpha_{e_k}(-e_k)$ such that the monopole lattice translations $S^\alpha_k$ defined by \eqref{eq-SalphaDef} satisfy
\begin{equation}
\label{eq-MonopoleTransConj}
\DiracPhase \,S^\alpha_k\,\DiracPhase 
\;=\;
S^{1-\alpha}_k
\;,
\end{equation}
$S^0_k=S_k$ and are continuous in $\alpha\in[0,1]$.  Furthermore, for even $d$
\begin{equation}
\label{eq-MonopoleTransConjEven}
\Gamma\,S^\alpha_k\,\Gamma
\;=\;
S^{\alpha}_k
\;.
\end{equation}
The operators $S^\alpha_k - S_k$ are compact on $\ell^2(\ZM^d,\CM^{2N})$.
\end{proposi}

\noindent {\bf Proof.} As already pointed out, both \eqref{eq-MonopoleTransConj} and \eqref{eq-MonopoleTransConjEven} hold for $S^\alpha_k$ away from $0$ and $-e_k$ by Proposition~\ref{prop-MonTrans}. At these points, \eqref{eq-MonopoleTransConj} becomes explicitly
\begin{equation}
\label{eq-DiracPhase2Points}
\DiracPhase(0)
\;=\;
M^{1-\alpha}_{e_k}(0)\DiracPhase(e_k)M^\alpha_{e_k}(0)^*
\;,
\qquad
\DiracPhase(0)
\;=\;
M^{\alpha}_{e_k}(-e_k)^*\DiracPhase(-e_k)M^{1-\alpha}_{e_k}(-e_k)
\;.
\end{equation}
Now from \eqref{eq-DGdef}, one has $\DiracPhase(e_k)=\gamma_k=-\DiracPhase(-e_k)$. At $\alpha=\frac{1}{2}$ these equations only involve $\DiracPhase(0)$, $M^{\frac{1}{2}}_{e_k}(0)$ and $M^{\frac{1}{2}}_{e_k}(-e_k)$. One can first choose $\DiracPhase(0)=\gamma_1$ (note that $\DiracPhase(0)=\one$ is not a possible choice) and then  $M^{\frac{1}{2}}_{e_k}(0)$ and $M^{\frac{1}{2}}_{e_k}(-e_k)$. Then one chooses paths $\alpha\in[0,\frac{1}{2}]\mapsto M^{\alpha}_{e_k}(0)$ and $\alpha\in[0,\frac{1}{2}]\mapsto M^{\alpha}_{e_k}(-e_k)$ connecting $\one$ to the  $M^{\frac{1}{2}}_{e_k}(0)$ and $M^{\frac{1}{2}}_{e_k}(-e_k)$ respectively. For even $d$, this can be done such that also \eqref{eq-MonopoleTransConjEven} holds. Finally, the equations \eqref{eq-DiracPhase2Points} can be used to extend these paths to $\alpha\in[\frac{1}{2},1]$. The last claim follows immediately from $S^\alpha_k - S_k=(M^\alpha_{e_k}-\one)S_k$ and the asymptotics \eqref{eq-MagTransAsymp}.
\hfill $\Box$



\section{Spectral flow of Hamiltonians with inserted monopole}
\label{sec-SFHam}

We consider a Hamiltonian on $\ell^2(\ZM^d,\CM^{n})$ of the form
\begin{equation}
\label{eq-HamFormBare}
h
\;=\;
\delta(S_1,\ldots,S_d)
\;+\;w
\;,
\end{equation}
where $\delta$ is a non-commutative polynomial with matrix coefficients specifying the kinetic part and $w=w^*$ is a matrix-valued potential. It is possible that both the coefficients of $\delta$ and $w$ are space dependent. The main hypothesis is a gap condition at the Fermi level $\mu\in\RM$, namely
\begin{equation}
\label{eq-GapHam}
\mu\,\not\in\,\sigma(h)
\;.
\end{equation}
Next let us tensor a representation space $\CM^{2N}$ for the Clifford algebra and then extend the Hamiltonian to 
\begin{equation}
\label{eq-HamFormBare2}
H=h\otimes\one_{2N}
\;.
\end{equation}
This is the procedure applied in the proof of the index theorems in \cite{PLB,PS,PS2}.  Note that $\sigma(h)=\sigma(H)$ and thus, in particular, $\mu\,\not\in\,\sigma(H)$. Now a monopole is inserted into $H$ by replacing $S_j$ by $S_j^\alpha$:
$$
H_\alpha
\;=\;
\delta(S^\alpha_1,\ldots,S^\alpha_d)
\;+\;W
\;,
$$
where $W=w\otimes \one_{2N}$. It follows from \eqref{eq-MonopoleTransConj} that
\begin{equation}
\label{eq-FHam}
\DiracPhase \,H_\alpha\,\DiracPhase 
\;=\;
H_{1-\alpha}
\;.
\end{equation}
Moreover, Proposition~\ref{prop-SalphaDef} implies that $\alpha\in[0,1]\mapsto H_\alpha-H_0$ is a compact selfadjoint operator. Hence there is an associated spectral flow through $\mu$ but it is not interesting: 

\begin{proposi}
\label{prop-NoSpecFlow}
One has:\quad
$
\SF
\big(\alpha\in[0,1]\mapsto H_\alpha\;\mbox{\rm by }\mu\big)
\;=\;
0
\;.
$
\end{proposi}

\noindent {\bf Proof.} The unitary equivalence \eqref{eq-FHam} implies for the spectra
$
\sigma(H_\alpha)
\;=\;
\sigma(H_{1-\alpha})
\;.
$
Due to this spectral symmetry there is thus no nett spectral flow.
\hfill $\Box$

\vspace{.15cm}

Nevertheless, the path $\alpha\in[0,1]\mapsto H_\alpha$ can have interesting topology if there is some supplementary symmetry. In this section the case of even $d$ is considered so that $\Gamma H_\alpha\Gamma=H_\alpha$ and thus $H_\alpha=\diag(h_\alpha,\tilde{h}_\alpha)$ with $\tilde{h}_0=h_0=h$.  The insertion of the monopole into $H=H_0$ then \eqref{eq-FHam} leads to
\begin{equation}
\label{eq-HamDiag}
{H}_\alpha
\;=\;
\begin{pmatrix}
h_\alpha & 0 \\ 0 & V\,h_{1-\alpha}\,V^*
\end{pmatrix}
\;,
\end{equation}
where $V$ is the unitary in the Dirac phase  \eqref{eq-GammaSpec}. For $d=2$, the formulas in Section~\ref{sec-Monopole234D} show that ${H}_\alpha=\diag(h_\alpha,h_{-\alpha})$ so that \eqref{eq-HamDiag} is indeed equivalent to the relation $h_{\alpha+1}=\DiracPhaseEntry^*h_\alpha \DiracPhaseEntry $ appearing in item (ii) in the introduction, as well as in \cite{DS}. It is also obvious in this case that the spectral flow in Proposition~\ref{prop-NoSpecFlow} decomposes into a direct sum of two spectral flows which cancel each other out. However, the remarkable point is that each of these spectral flows is defined by itself. 

\begin{theo}
\label{theo-SpecFlow}
Let $d$ be even and suppose $h$ is of the form \eqref{eq-HamFormBare}. For $\mu\not\in\sigma(h)$, one then has
$$
\SF
\big(\alpha\in[0,1]\mapsto h_\alpha\;\mbox{\rm by }\mu\big)
\;=\;
\Ind(p_\mu \DiracPhaseEntry  p_\mu)
\;,
$$
where $p_\mu=\chi(h\leq \mu)$ is the spectral projection of $h$ on states below $\mu$.
\end{theo}

\noindent {\bf Proof.} Due to $h_1=V^*\tilde{h}_0V=V^*h_0V$, this immediately follows from Phillip's result \cite{Ph1} stated as Theorem~\ref{theo-SFInd1} in the appendix.
\hfill $\Box$

\vspace{.15cm}

For $d=2$, Theorem~\ref{theo-SpecFlow}  reproduces the result of \cite{DS} described in the introduction. Let us next provide an explicit instance in higher dimension where Theorem~\ref{theo-SpecFlow} applies and leads to a non-trivial spectral flow.

\vspace{.15cm}

\noindent {\bf Example} We follow Section~2.2.4 of \cite{PS2}. For even $d$, let $\nu_1,\ldots,\nu_d$ be an irreducible representation of $\CM_d$ on $\CM^{n}$ with grading $\nu_0$. Consider the Hamiltonian on $\ell^2(\ZM^d,\CM^{n})$ given by
$$
h\;=\;
\frac{1}{2\imath}\sum_{j=1}^d(S_j-S_j^*)\nu_j
\;+\;
\left(
m+\frac{1}{2}\sum_{j=1}^d(S_j+S_j^*)
\right)
\nu_0
\;.
$$
Then $0\not\in\sigma(h)$ for $m\not\in \{-d,-d+2,\ldots,d\}$. At these values the central gap at the Fermi level $\mu=0$ closes, and actually the $d$-th Chern number $\Ch_d(p)$ of the Fermi projection $p=\chi(h\leq 0)$ changes its value. The reader is referred to \cite{PS2} for an explicit calculation of  all the values taken. Now the second Clifford representation $\gamma_1,\ldots,\gamma_d$  on $\CM^{2N}$ is tensorized to the Hilbert space so that the new Hamiltonian $H=h\otimes \one_{2N} $  and the Dirac operator $D$, identified with $D\otimes \one_n$, act on the same Hilbert space $\ell^2(\ZM^d,\CM^{2Nn})$. In this situation, Chapter~6 of \cite{PS2} shows that $\Ch_d(p)=\Ind(p_\mu \DiracPhaseEntry  p_\mu)$. Thus the index appearing in Theorem~\ref{theo-SpecFlow} takes non-trivial values. Consequently, the spectral flow of $\alpha\in[0,1]\mapsto h_\alpha$ past $0$ is non-trivial as well.
\hfill $\diamond$

\section{Chirality flow in odd dimensions}
\label{sec-ChirFlow}

Let us begin by illustrating the claims (i)$^\prime$ to (iii)$^\prime$ stated in Section \ref{sec-ChiralFlow} on a simple one-dimensional toy model, the Su-Schrieffer-Heeger model \cite{SSH} with vanishing mass and no disorder. The Hamiltonian with inserted flux $\alpha$ in one cell of the $2$-strip acts on $\ell^2(\ZM,\CM^2)$. It is 
\begin{equation}
\label{eq-SIntro}
H_\alpha
\;=\;
\begin{pmatrix}
0 & S^\alpha
\\
(S^\alpha)^* & 0
\end{pmatrix}
\;,
\end{equation}
where the operator $S^\alpha$ is the bilateral shift operator perturbed by a rank one operator depending on $\alpha$. Using Dirac's bra-ket notations, one has:

\begin{equation}
\qquad
S^\alpha
\;=\;
\sum_{n\not=0}|n\rangle\langle n+1|
\;+\;
e^{\imath \pi\alpha}|0\rangle\langle 1|
\;,
\end{equation}
so that the Fermi unitary is $U_\alpha=S^\alpha$. The Dirac phase $\DiracPhase $ is the sign of the position operator, namely
\begin{equation}
\label{eq-FIntro}
\DiracPhase 
\;=\;
\sum_{n>0}|n\rangle\langle n|
\;-\;
\sum_{n\leq 0}|n\rangle\langle n|
\;.
\end{equation}
One readily checks that all items (i)$^\prime$ to (iii)$^\prime$ hold. In particular,
$$
\DiracPhase S^\alpha (S^0)^*
\;=\;
\sum_{n>0}|n\rangle\langle n|
\;-\;
e^{\imath \pi\alpha}|0\rangle\langle 0|
\;-\;
\sum_{n<0}|n\rangle\langle n|
\;,
$$
so that $\SF\big(\alpha\in[0,1]\mapsto \DiracPhase S^\alpha (S^0)^*\big)=1=\Ind(\Hardy S^0\Hardy)$. While this results from a direct elementary calculation, the homotopy invariance of both quantities in the equality in (iii)$^\prime$ allows one to extend it to a model with a random mass term, as discussed in \cite{MSHP,PS2}. 

\vspace{.15cm}

In the remainder of this section, the space dimension $d\geq 3$ is odd. Furthermore, the Hamiltonian is of the form \eqref{eq-HamFormBare2}. Moreover, it is supposed to have a local translation invariant chiral symmetry. This requires $n=2n'$ to be even. Then the chiral symmetry operator on $\ell^2(\ZM^d,\CM^{2Nn})$ is
$$
J
\;=\;
\begin{pmatrix} 
\one_{n'} & 0 \\ 0 & -\,\one_{n'}
\end{pmatrix}
\otimes\one_{2N}
\;.
$$
The Hamiltonian on $\ell^2(\ZM^d,\CM^{2Nn})$ then has the chiral symmetry
$
J\,H\,J\;=\;-H
\;.
$
This implies the off-diagonal form  
$
H
\;=\;
\begin{pmatrix}
0 & B \\ B^* & 0
\end{pmatrix}
\;,
$
with an operator $B$ on $\ell^2(\ZM^d,\CM^{n'})\otimes\CM^{2N}$ acting trivially (as identity) on the fiber $\CM^{2N}$. Again $B$ is the sum of a matrix-valued polynomial in the shift operators $S_1,\ldots,S_d$ and a matrix potential. One can hence replace the shift operators $S_j$ by the unitary monopole shifts $S^\alpha_j$ to obtain operators $B_\alpha$ and a family of chiral Hamiltonians $H_\alpha$:
$$
H_\alpha
\;=\;
\begin{pmatrix}
0 & B_\alpha \\ B_\alpha^* & 0
\end{pmatrix}
\;.
$$
By Proposition~\ref{prop-SalphaDef} the differences $H_\alpha-H_0$ and $B_\alpha-B_0$ are compact. The Fermi level for chiral Hamiltonians is $\mu=0$ and the gap hypothesis \eqref{eq-GapHam} states that $H_0=H$ is invertible, and hence also $B=B_0$ is invertible. Moreover, the path $\alpha\in[0,1]\mapsto H_\alpha$ is real analytic. Therefore Proposition~\ref{prop-ChirInd}, stated in the appendix, implies the following:

\begin{theo}
\label{theo-ChirFlow}
Let $d$ be odd and suppose that the chiral Hamiltonian $H$ is of the form \eqref{eq-HamFormBare2} and $\mu=0$. Then
$$
\SF(\alpha\in[0,1]\mapsto J\DiracPhase H_\alpha H_0^{-1})
\;=\;2\;\Ind(\Hardy B_0\Hardy)
\;, 
$$
where $\Hardy=\frac{1}{2}(\DiracPhase +\one)$ is the Hardy projection of $\DiracPhase $.
\end{theo}

Let us note again that the spectral flow in the theorem results from a path connecting $J\DiracPhase $ to $-H_0J\DiracPhase  H_0^{-1}$ that we also refer to  as the chirality flow. If the whole path $\alpha\in[0,1]\mapsto H_\alpha$ is invertible, one can define the path of Fermi unitaries $\alpha\in[0,1]\mapsto U_\alpha$ by
$$
H_\alpha\,|H_\alpha|^{-1}
\;=\;
\begin{pmatrix}
0 & U_\alpha^* \\ U_\alpha & 0
\end{pmatrix}
\;,
$$
and Theorem~\ref{theo-ChirFlow} reduces to item (iii)$^\prime$ in the introduction. This is the case in the SSH model discussed in the introduction.

\vspace{.15cm}

\noindent {\bf Example}.  Section~2.3.3 of \cite{PS2} provides for odd $d$ an example of a chiral model with non-vanishing strong invariant. As this is very similar to the even dimensional model presented in Section~\ref{sec-SFHam}, no further details are spelled out. 
\hfill $\diamond$

\appendix

\section{Spectral flow between unitary conjugates}
\label{sec-SF}

This appendix elaborates on a result of Phillips \cite{Ph1} which connects the spectral flow between unitary conjugate selfadjoint Fredholm operators to the index of a Fredholm operator.  Intuitively, the spectral flow of a given path $\alpha\in[0,1]\mapsto T_\alpha$ of selfadjoint Fredholm operators on a Hilbert space $\Hh$ simply counts the number of eigenvalues passing by the origin, weighted with the orientation of the passage. We will not review the technical details of Phillips' analytic approach \cite{Phi} nor the refinements from \cite{BCPRSW} as they are not needed in the following, and simply refer the interested reader to \cite{Phi,Ph1,BCPRSW}. Let us just note that the spectral flow is a homotopy invariant under deformation of the path if the end points are fixed. The following result of Phillips \cite{Ph1} is used in this paper:

\begin{theo}
\label{theo-SFInd1}
Let $\alpha\in[0,1]\mapsto T_\alpha$ be a path of selfadjoint Fredholm operators with invertible end points $T_0$ and $T_1$ and such that $T_\alpha-T_0$ is compact. Furthermore, suppose that there exists a unitary $U$ such that 
$
T_1\;=\;
U^*T_0U
\;.
$
If $P=\chi(T_0\leq 0)$, then $PUP$ is a Fredholm operator on $\Ran(P)$ with index given by \:
$
\Ind(PUP)
\;=\;
\SF\big(
\alpha\in[0,1]\mapsto T_\alpha
\big)
\;.
$
\end{theo}

The proof in \cite{Ph1} is based on the index of Fredholm pairs of projections \cite{ASS}. A proof by a homotopy argument is given in \cite{DS2}. This section proves a unitary equivalent of this result (based on the formulas on pp.~60-61 in \cite{DS2}). For this purpose, the notion of spectral flow is first slightly extended to also allow a path $\alpha\in[0,1]\mapsto T_\alpha$ of normal operators with self-adjoint and invertible end points $T_0=T_0^*$ and $T_1=T_1^*$ such that $T_\alpha-T_0$ is compact. In this situation, the definition of the spectral flow using spectral projections as given in \cite{Phi} immediately transposes (one just considers spectral projections associated to eigenvalues lying in vertical strips of the complex plane). Alternatively, one can simply use the spectral flow of self-adjoint Fredholm operators, namely
\begin{equation}
\label{eq-SFnormal}
\SF\big(
\alpha\in[0,1]\mapsto T_\alpha
\big)
\;=\;
\SF\big(
\alpha\in[0,1]\mapsto \Re e(T_\alpha)
\big)
\;,
\end{equation}
where $\Re e(T_\alpha)=\frac{1}{2}(T_\alpha+T_\alpha^*)$ is a selfadjoint Fredholm operator. For this type of spectral flow one now has an analogue to Theorem~\ref{theo-SFInd1}.

\begin{theo}
\label{theo-SFInd2}
Let $\alpha\in[0,1]\mapsto U_\alpha$ be a path of unitaries such that $U_\alpha- U_0$ is compact. Suppose that there is a selfadjoint unitary $\DiracPhase $ such that
$
U_1\;=\;\DiracPhase U_0\DiracPhase 
\;.
$
If $\Hardy=\chi(\DiracPhase \geq 0)$, then $\Hardy U_0\Hardy$ is a Fredholm operator on $\Ran(\Hardy)$ with index given by
\begin{equation}
\Ind(\Hardy U_0\Hardy)
\;=\;
\SF\big(
\alpha\in[0,1]\mapsto W_\alpha
\big)
\;,
\label{eq-UnitaryInd}
\end{equation}
where 
$
W_\alpha\;=\;\DiracPhase U_\alpha U_0^*
\;.
$
\end{theo}

\noindent {\bf Proof.} To show that the spectral flow is well-defined, we note that $U_\alpha U_0^*-\one=(U_\alpha-U_0)U_0^*$ is compact by hypothesis, so that $W_\alpha-\DiracPhase $ is compact  and so is $\Re e(W_\alpha)-\DiracPhase $. Moreover, $W_0=\DiracPhase =\Re e(W_0)$ and $W_1=U_0\DiracPhase U_0^*=\Re e(W_1)$ are both self-adjoint, and $\Re e(W_1)=U_0 \Re e(W_0)U_0^*$. Hence Theorem~\ref{theo-SFInd1} can be applied to the family $\alpha\in[0,1]\mapsto T_\alpha=\Re e(W_\alpha)$.  There are now two sign changes in the index pairing involved, one because $\Hardy$ is the spectral projection onto the positive spectrum of $\DiracPhase$ and one because $U_0$ is on the l.h.s. in $\Re e(W_1)=U_0 \Re e(W_0)U_0^*$ (while $P$ is the negative spectral projection of $T_0$ and $T_1=
U^*T_0U$ in Theorem~\ref{theo-SFInd1}). This concludes the proof.
\hfill $\Box$

\vspace{.15cm}

Let us stress that the spectral flow of unitaries in \eqref{eq-UnitaryInd} does {\em not} distinguish whether the eigenvalue travels on the upper or lower half of the unit circle, in contradistinction to the spectral flow of essentially gapped unitaries \cite{KL,SB}. The spectral flow of unitaries has the same homotopy invariance properties as the spectral flow of the paths of selfadjoints. For example, choosing $U_0U_\alpha^* \DiracPhase $ instead of $W_\alpha$ is another natural choice giving a different path connecting $\DiracPhase $ and $U_0\DiracPhase U_0^*$. The choices $\DiracPhase U_\alpha^*U_0$  and $U_0^*U_\alpha \DiracPhase $ reverse the path and thus the sign of the spectral flow. A standard form of a path from $\DiracPhase $ to $U_0\DiracPhase U_0^*$, expressed merely in terms of $U_0$ and $\DiracPhase $, is 
$$
U_\alpha
\;=\;
U_0\exp\big(\tfrac{\imath\pi}{2}(\DiracPhase -\one+\alpha\,U_0^*[\DiracPhase ,U_0])
\big)
\;.
$$
This was already given in \cite{DS2} where it is also shown that this path establishes an isomorphism between two $K_1$-groups, one of a C$^*$-algebra containing $U$ and one of the associated mapping cone. Theorem~\ref{theo-SFInd2} evaluates this $K$-theoretic fact, just as Theorem~\ref{theo-SFInd1} results from an isomorphism of the $K_0$-group of the algebra to the $K_1$-group of an ideal in the mapping cone \cite{DS2}. At this point, all relevant preparations for the main text (in particular, Theorem~\ref{theo-ChirFlow}) are attained. What follows are  further comments and auxiliary results on the spectral flow defined by \eqref{eq-SFnormal}. 

 
\vspace{.15cm}

From the above $K$-theoretic perspective, it is natural to also consider paths  $\alpha\in[0,1]\mapsto B_\alpha$ of invertible operators with compact differences $B_\alpha-B_0$ and $B_1=\DiracPhase B_0\DiracPhase $ where $\DiracPhase $ is a selfadjoint unitary as above. Then consider the path $\alpha\in[0,1]\mapsto T_\alpha=\DiracPhase  B_\alpha B_0^{-1}$ of invertibles connecting the selfadjoint unitary $T_0=\DiracPhase $ to the operator $T_1=B_0\DiracPhase B_0^{-1}$ which also has spectrum $\{-1,1\}$. For this path, one would like to define a spectral flow. If the operators are normal, the above procedure works. For a more general (not necessarily normal) case, let us suppose that the path $\alpha\in[0,1]\mapsto T_\alpha$ is real analytic and  (merely) assume that
$
T_\alpha-T_0\;\;\mbox{is compact with }\;T_0\,,\;T_1\;\;\mbox{invertible with real spectrum}\;.
$
As the unbounded component of the resolvent set of $T_0$ contains $\CM\setminus\RM$, it then follows  from analytic Fredholm theory that the essential spectrum of $T_\alpha$ coincides with the essential spectrum of $T_0$. Hence by analytic perturbation theory \cite{Kat} the discrete eigenvalues of $T_\alpha$ vary analytically away from level-crossings, at which there may be root singularities (Puiseux expansion). Therefore the spectral flow through the imaginary axis along the path $\alpha\in[0,1]\mapsto T_\alpha$ is well-defined by counting the finite number of eigenvalues passing from the left half-plane to the right half-plane, minus those from passing right to left.  In general, this spectral flow is {\em not} given by the r.h.s. of \eqref{eq-SFnormal}. Indeed, if say $T_1$ is not normal, the spectra of $T_1$ and of $\Re e(T_1)$ may have little in common. The generalized spectral flow will still be denoted by $\SF\big(\alpha\in[0,1]\mapsto T_\alpha\big)$. It is invariant under analytic homotopies of the path provided the above conditions are satisfied. This allows us to define the spectral flow for an analytic path $\alpha\in[0,1]\mapsto B_\alpha$ as above. 

\vspace{.15cm}

A special case is the set-up in Theorem~\ref{theo-SFInd2} in which $B_\alpha$ is unitary. To put ourselves in this situation, let us use the real analytic homotopy of paths:
$
s\in[0,1]\;\mapsto\;B_{\alpha,s}\,=\,B_\alpha\,|B_\alpha|^{-s}
\;.
$
Then $B_{\alpha,1}$ is indeed unitary so that Theorem~\ref{theo-SFInd2} applies. Furthermore, both quantities in \eqref{eq-UnitaryInd} are homotopy invariant. Moreover, $2[\Hardy,B_0]=[\DiracPhase ,B_0]=(\DiracPhase B_0\DiracPhase -B_0)\DiracPhase =(B_1-B_0)\DiracPhase $ is compact so that $\Hardy B_0\Hardy$ is Fredholm on $\Ran(\Hardy)$ because $\Hardy B_0^{-1}\Hardy$ is a pseudo inverse. Finally, by standard arguments $s\mapsto\Hardy B_{0,s}\Hardy$ is a continuous path of Fredholm operators which hence have constant index.  In conclusion, for an analytic part of invertibles with end points that are conjugate by a selfadjoint unitary $\DiracPhase =(\one-\Hardy)-\Hardy$,
\begin{equation}
\label{eq-IndNonNormal}
\Ind(\Hardy B_0\Hardy)
\;=\;
\SF\big(
\alpha\in[0,1]\mapsto \DiracPhase  B_\alpha B_0^{-1}
\big)
\;.
\end{equation}
Let us note that the spectral flow on the r.h.s. is well-defined if merely the initial point $B_0$ is invertible. This will be explored further towards the end of this section.

\vspace{.15cm}

The next series of comments shows that by passing to $2\times 2$ matrices one can rewrite Theorem~\ref{theo-SFInd2} as a version of Theorem~\ref{theo-SFInd1} with a supplementary symmetry. Indeed, given the situation of Theorem~\ref{theo-SFInd2}, let us set
\begin{equation}
\label{eq-PRep}
P_\alpha
\;=\;
\frac{1}{2}
\begin{pmatrix}
\one & -\,U_\alpha^* \\ -\,U_\alpha & \one
\end{pmatrix}
\;,
\qquad
J\;=\;
\begin{pmatrix}
\one & 0 \\ 0 & -\one
\end{pmatrix}
\;.
\end{equation}
Let us also extend $\DiracPhase $ by identifying it (by abuse of notation) with $\DiracPhase \otimes \one_2=\diag(\DiracPhase ,\DiracPhase )$. Then $J\DiracPhase =\DiracPhase J=\diag(\DiracPhase ,-\DiracPhase )$ is also a selfadjoint unitary with spectrum $\{-1,1\}$. All these operators act on $\Hh\otimes\CM^2$ which becomes a Krein space with fundamental symmetry $J$ (which plays the role of the chiral symmetry operator in the application in Section~\ref{sec-ChirFlow}). Recall, {\it e.g.} from \cite{KL,SB} and references therein, that a Krein space is a Hilbert space equipped with a selfajoint unitary $J$ called the fundamental symmetry or fundamental form as it is often also viewed as a sesquilinear form $(v,w)\in\Hh\times\Hh\mapsto\langle v|Jw\rangle$. On such a Krein space, one then has the notions of $J$-isotropic subspace, namely a subspace on which the form $J$ vanishes. Maximal $J$-isotropic subspaces are also $J$-Lagrangian. Orthogonal projections on such subspaces are also called $J$-isotropic or $J$-Lagrangian. The projection $P_\alpha$ is $J$-Lagrangian, namely it satisfies
$$
J\,P_\alpha\,J
\;=\;
\one-P_\alpha
\;.
$$
Let us recall from \cite{KL,SB} some further standard facts on $J$-Lagrangian projections on a Krein space. Clearly, the set of these $J$-Lagrangian projections can be identified with the set of closed $J$-Lagrangian subspaces, also called the Lagrangian Grassmannian. Moreover, the $J$-Lagrangian projections are always of the form \eqref{eq-PRep} and are thus in bijection with the set of unitary operators from the positive onto the negative eigenspaces of $J$. In our particular situation, $P_\alpha-P_0$ is compact and this implies that $P_\alpha$ and $JP_0$ form a Fredholm pair. This means that $P_\alpha$ and $JP_0$ are essentially transversal, in the sense that the intersection of the ranges and co-ranges are finite dimensional and the essential angle spectrum between the two ranges does not contain $0$. In this situation, it is natural to consider the Bott-Maslov index counting the weighted number (by orientation) of intersections of $P_\alpha$ with the singular cycle $JP_0$. This is given by the oriented spectral flow of the path $\alpha\in[0,1]\mapsto U_0^*U_\alpha$ through $-1$ (counter clockwise passages give positive contributions) \cite{KL,SB}. This path has end points $\one$ and $U_0^*\DiracPhase U_0\DiracPhase $. Thus the Bott-Maslov index is {\em not} invariant under  homotopic deformations of either $U_0$ or $\DiracPhase $. It is not linked to spectral flow in Theorem~\ref{theo-SFInd2}. We rather consider the path
$$
\alpha\in[0,1]
\;\mapsto\;
T_\alpha
\;=\;
(\one-P_\alpha)\,-\,P_\alpha
\;=\;
\begin{pmatrix}
0 & U_\alpha^* \\ U_\alpha & 0
\end{pmatrix}
\;,
$$
of selfadjoint unitary Fredholm operators. It satisfies $T_1=\DiracPhase T_0\DiracPhase $ as in Theorem~\ref{theo-SFInd1}, but, moreover, it has the chiral symmetry $JT_\alpha J=-T_\alpha$. The spectral flow of $\alpha\in[0,1]\mapsto T_\alpha$ vanishes so that Theorem~\ref{theo-SFInd1} is trivial in this situation, and one has to proceed as in Theorem~\ref{theo-SFInd2} and extract topological information from the path $\alpha\in[0,1]\mapsto U_\alpha$.  The following proposition now considers a more general path of chiral selfadjoint Fredholm operators. This path need not consist of  unitaries and indeed need not even consist of invertibles.

\begin{proposi}
\label{prop-ChirInd}
Let $\alpha\in[0,1]\mapsto T_\alpha$ be an analytic path of selfadjoint Fredholm operators on the Krein space $\big(\Hh\otimes\CM^2,J=\diag(\one,-\one)\big)$ satisfying the chiral symmetry $JT_\alpha J=-T_\alpha$, as well as  $T_1=\DiracPhase T_0\DiracPhase $ with an invertible $T_0$ and a selfajoint unitary $\DiracPhase =\DiracPhase \otimes \one_2$. Due to the chiral symmetry, there is a path $\alpha\in[0,1]\mapsto B_\alpha$ of operators such that
$$ 
T_\alpha\;=\; \begin{pmatrix} 0 & B_\alpha^* \\ B_\alpha & 0 \end{pmatrix}\;, 
$$
Then $\Hardy B_0\Hardy$ with $\Hardy=\frac{1}{2}(\DiracPhase +\one)$ is a Fredholm operator on $\Ran(\Hardy)$ with
$$
2\;\Ind(\Hardy B_0\Hardy)
\;=\;
\SF\big(
\alpha\in[0,1]\mapsto J\DiracPhase T_\alpha T_0^{-1}
\big)
\;.
$$
\end{proposi}

\noindent {\bf Proof.} Set $C_\alpha=J\DiracPhase T_\alpha T_0^{-1}$. First of all, the path $\alpha\in[0,1]\mapsto C_\alpha$ connects $J\DiracPhase $ with $- T_0J\DiracPhase T_0^{-1}$ which both have spectrum $\{-1,1\}$. Moreover, $C_\alpha-J\DiracPhase =J\DiracPhase (T_\alpha-T_0) T_0^{-1}$ is compact so that, as above, the spectral flow of  the analytic path $\alpha\in[0,1]\mapsto C_\alpha$ is well-defined. Now $B_\alpha-B_0$ is compact and $B_1=\DiracPhase B_0\DiracPhase $. One has
$$
C_\alpha
\;=\;
\begin{pmatrix} \DiracPhase  B_\alpha B_0^{-1} & 0 \\ 0 & -(B_0^{-1}B_\alpha \DiracPhase )^*  \end{pmatrix}
\;.
$$
In the  case that all $T_\alpha$ are invertible, so too are all $B_\alpha$ and then the spectral flow of $C_\alpha$ is the direct sum of two contributions which are both equal to the spectral flow on the r.h.s. of \eqref{eq-IndNonNormal}. This implies the result in this case. In the other case, the invertibility of $T_\alpha$ can only fail on a finite number of points. In each such point, the kernel is even dimensional and the restriction of $J$ to this kernel has vanishing signature. A generic finite dimensional perturbation on the kernel will render the operator $T_\alpha$ invertible. We will not carry this out in detail, but provide an example below.
\hfill $\Box$

\vspace{.15cm}

\noindent {\bf Example}. Let us consider the $S^0$ and $S^1$ on $\ell^2(\ZM)$ as given in \eqref{eq-SIntro}. Hence $S^1=\DiracPhase S^0\DiracPhase $ with $\DiracPhase $ as in \eqref{eq-FIntro}. Instead of the path $\alpha\in[0,1]\mapsto H_\alpha$ of chiral selfadjoint unitaries  given in \eqref{eq-SIntro}, let us consider another path connecting $H_0$ and $H_1$:
$$
\alpha\in[0,1]\;\mapsto\;T_\alpha
\;=\;
\begin{pmatrix}
0 & S^0 - 2\alpha\,|0\rangle\langle 1| \\
(S^0 - 2\alpha\,|0\rangle\langle 1|)^* & 0
\end{pmatrix}
\;.
$$
Clearly this path consists again of chiral selfadjoint Fredholm operators, but for $\alpha=\frac{1}{2}$ the invertibility is not given. However, replacing $\alpha\in[0,1]\mapsto (1-2\alpha)$ in the definition of $T_\alpha$ by any path from $1$ to $-1$ avoiding $0$, leads to the invertibility of the whole path.
\hfill $\diamond$

\vspace{.15cm}

\noindent {\bf Acknowledgements:} We thank the referees and Nora Doll for several constructive comments on the first draft of this paper. The work of A.~L.~C. was supported by the Australian Research Council, that of H.~S.-B. partially by the DFG. 


\end{document}